\documentclass[twocolumn]{aastex701}

\usepackage{footnote}
\usepackage{graphicx}	
\usepackage{amsmath}	
\usepackage{newtxtext,newtxmath} 
\usepackage{multirow} 
\usepackage{lineno}

\newcommand{\Msun}{M$_{\odot}$}
\newcommand{\Lsun}{L$_{\odot}$}
\newcommand{\Mbh}{$M_{\rm BH}$}

\newcommand{\ml}{\emph{M/L}}

\newcommand{\hst}{\emph{HST}}
\newcommand{\jwst}{\emph{JWST}}
\newcommand{\kms}{km~s$^{-1}$}

\newcommand{\go}{G$_{\rm outer}$}

\newcommand{\jamcyl}{JAM\textsubscript{cyl}}
\newcommand{\jamsph}{JAM\textsubscript{sph}}

\hypersetup{colorlinks, linkcolor=blue, citecolor=cyan, urlcolor=magenta}

\received{September 23, 2025} 
\revised{December 10, 2025}
\accepted{to The Astrophysical Journal \today}

\shorttitle{Measuring SMBHs in NGC 4258 with JWST/NIRSpec}
\shortauthors{D. D.\ Nguyen, H.\ N.\ Ngo $\&$ M. Cappellari et al.}
\begin{document}

\title{Measuring the Central Dark Mass in NGC 4258 with JWST/NIRSpec Stellar Kinematics}

\correspondingauthor{Dieu D.\ Nguyen}\email{dieun@umich.edu}
\author[0000-0002-5678-1008]{Dieu D.\ Nguyen}
\email{dieun@umich.edu}
\affil{Department of Astronomy, University of Michigan, 1085 South University Avenue, Ann Arbor, MI 48109, USA}

\author[0009-0006-5852-4538]{Hai N.\ Ngo}
\email{hai10hoalk@gmail.com}
\affiliation{Faculty of Physics – Engineering Physics, University of Science, Vietnam National University in Ho Chi Minh City, Vietnam}

\author[0000-0002-1283-8420]{Michele Cappellari}
\email{michele.cappellari@physics.ox.ac.uk}
\affiliation{Sub-Department of Astrophysics, Department of Physics, University of Oxford, Denys Wilkinson Building, Keble Road, Oxford, OX1 3RH, UK}

\author[0009-0004-3689-8577]{Tinh Q.\  T.\ Le}
\email{lethongquoctinh01@gmail.com}
\affiliation{Department of Physics, International University, Vietnam National University in Ho Chi Minh City, Vietnam}

\author[0009-0005-8845-9725]{Tien  H.\  T.\ Ho}  
\email{htien2808@gmail.com}
\affiliation{Faculty of Physics – Engineering Physics, University of Science, Vietnam National University in Ho Chi Minh City, Vietnam}

\author[0009-0005-8845-9725]{Tuan N.\ Le}
\email{htien2808@gmail.com}
\affiliation{International Centre for Interdisciplinary Science and Education, 07 Science Avenue, Ghenh Rang, 55121 Quy Nhon, Vietnam}

\author[0000-0001-5802-6041]{Elena Gallo}
\email{egallo@umich.edu}
\affil{Department of Astronomy, University of Michigan, 1085 South University Avenue, Ann Arbor, MI 48109, USA}

\author[0000-0002-6694-5184]{Niranjan Thatte} 
\email{niranjan.thatte@physics.ox.ac.uk}
\affiliation{Sub-Department of Astrophysics, Department of Physics, University of Oxford, Denys Wilkinson Building, Keble Road, Oxford, OX1 3RH, UK}

\author[0000-0002-4436-6923]{Fan Zou}
\email{fanzou@umich.edu}
\affil{Department of Astronomy, University of Michigan, 1085 South University Avenue, Ann Arbor, MI 48109, USA}

\author[0000-0002-0362-5941]{Michele Perna}
\email{fmperna@cab.inta-csic.es}
\affiliation{Centro de Astrobiolog\'{i}a (CAB), CSIC–INTA, Departamento de Astrof\'{i}sica, Cra. de Ajalvir Km. 4, 28850 – Torrej\'{o}n de Ardoz, Madrid, Spain}

\author[0000-0002-4005-9619]{Miguel Pereira-Santaella}
\email{miguel.pereira@iff.csic.es}
\affiliation{Instituto de F\'isica Fundamental, CSIC, Calle Serrano 123, 28006 Madrid, Spain}

\begin{abstract}

We present a new stellar dynamical measurement of the supermassive black hole (SMBH) mass in the nearby spiral galaxy NGC~4258 (M106), a critical benchmark for extragalactic mass measurements. We use archival \jwst/NIRSpec IFU data (G235H/F170LP grating) to extract high-resolution two-dimensional stellar kinematics from the CO bandhead absorption features within the central $3\arcsec \times 3\arcsec$. We extract the stellar kinematics after correcting for instrumental artifacts and separating the stellar light from the non-thermal AGN continuum. We employ Jeans Anisotropic Models (JAM) to fit the observed kinematics, exploring a grid of 12 models to systematically test the impact of different assumptions for the point-spread function, stellar mass-to-light ratio ($M/L$) profile, and orbital anisotropy. All 12 models provide broadly acceptable fits, albeit with minor differences. The ensemble median and  68\% (1$\sigma$) bootstrap confidence intervals of our 12 models yield a black hole mass of $M_{\rm BH} = (4.08^{+0.19}_{-0.33}) \times 10^7$~\Msun. This paper showcases the utility of using the full model ensemble to robustly account for systematic uncertainties, rather than relying on formal errors from a single preferred model, as has been common practice. Our result is just 5\% larger than, and consistent with, the benchmark SMBH mass derived from water maser dynamics, validating the use of NIRSpec stellar kinematics for robust SMBH mass determination. Our analysis demonstrates \jwst's capability to resolve the SMBH's sphere of influence and deliver precise dynamical masses, even in the presence of significant AGN continuum emission.
\end{abstract}



\section{Introduction}\label{sec:intro}

NGC~4258 (M106) is a nearby spiral galaxy of Hubble type SABbc, located at a distance $D=7.2 \pm 0.3$ Mpc, the most precise extragalactic distance measured \citep{Herrnstein1999}. It hosts one of the closest known active galactic nuclei (AGN) with a nuclear non-stellar continuum and broad optical emission lines detected in polarized light \citep{Wilkes1995}, consistent with the presence of an obscured central engine. This galaxy is particularly notable for containing one of the best-constrained supermassive black hole (SMBH) masses (\Mbh) in the nearby universe \citep{Miyoshi1995}. Using Very Long Baseline Interferometry (VLBI) observations of a rotating water-maser disk within 0.13 pc of the nucleus, a precise dynamical mass of $M_\mathrm{BH}^{\rm maser}$\ $=(3.9 \pm 0.1) \times 10^7~{\rm M}_\odot\ (D/7.2~\text{Mpc})$ was determined \citep{Herrnstein1999}. Aside from Sgr A* in the Milky Way, whose mass has been dynamically measured from the full orbits of individual stars via optical observations \citep{Genzel1996, Ghez1998}, NGC~4258 remains the most precisely measured extragalactic \Mbh\ to date. It serves as a critical benchmark for calibrating other \Mbh\ estimation methods.

Several previous studies have explored the SMBH in NGC~4258 using a variety of observational data and dynamical modeling approaches. These include: (i) refined maser dynamical modeling \citep{Herrnstein2005}; (ii) long-slit gas kinematics from the {\it Hubble Space Telescope} (\hst) Space Telescope Imaging Spectrograph (STIS), modeled under the assumption of circular motion in a thin ionized gas disk, yielding \Mbh$_{\rm , gas}$\ $=(7.9^{+2.5}_{-1.4}) \times 10^7$~\Msun\ \citep[converted to $1\sigma$ confidence]{Pastorini2007}; (iii) stellar kinematics based on \ion{Ca}{2} Triplet (CaT at $\sim$0.854 \micron) absorption lines from \hst/STIS and orbit-based modeling using the \citet{Schwarzschild1979} method, resulting in \Mbh$_{\rm , stars}$\ $=(3.3 \pm 0.2) \times 10^7$~\Msun\ \citep{Siopis2009}; and (iv) stellar kinematics derived from CO bandheads absorptions at $\sim$2.3 \micron\ using the Gemini Near-Infrared Integral-Field Spectrograph (NIFS) and modeled with Jeans Anisotropic Models \citep[JAM;][]{Cappellari2008, Cappellari2020}, giving \Mbh$_{\rm , stars}$\ $=(4.8\pm0.3) \times 10^7$~\Msun\ \citep[converted to $1\sigma$ confidence]{Drehmer2015}. These results are consistent with the maser-based estimate \citep{Miyoshi1995, Herrnstein2005} within their quoted uncertainties.

Given the SMBH mass inferred from maser disk dynamics and adopting a bulge velocity dispersion of $\sigma\approx105$~\kms\ \citep{Drehmer2015}, the SMBH's radius of the sphere of influence (SOI\footnote{The spherical region surrounding black hole, where its gravitational potential dominates and the enclosed stellar mass equals approximately twice the black hole mass, can be estimated as $r_{\rm SOI} \approx G M_{\rm BH}/\sigma^2$, where $G$ is the gravitational constant.}) is estimated as $r_{\rm SOI} \approx 15$ pc. Assuming a flat $\Lambda$CDM cosmology ($H_0 \approx 70$ km s$^{-1}$ Mpc$^{-1}$, $\Omega_{\rm m,0} \approx 0.3$, and $\Omega_{\rm \Lambda,0} \approx 0.7$), the corresponding angular scale is 35 pc $\arcsec^{-1}$, implying $r_{\rm SOI} \approx 0\farcs42$. This is two--three times larger than the point spread function (PSF) full width at half maximum of the {\it James Webb Space Telescope} (\jwst) Near-Infrared Spectrograph (NIRSpec), which has FWHM$_{\rm PSF} \approx  0\farcs15-0\farcs18$ at 2.4 \micron\ \citep[][]{D’Eugenio2024NaAs}. 
 
In this work, we analyzed the stellar kinematics of the nuclear region of NGC~4258 using archival \jwst/NIRSpec data (PID: 02016, PI: Anil C. Seth) in combination with JAM to constrain the dynamical properties of the system.  By employing the JAM framework \citep[e.g.,][]{Ahn2018, Davis2020, Nguyen17conf, Nguyen2017, Nguyen2018, Nguyen2019, Nguyen2025b}, we explore how variations in key model parameters, such as inclination ($i$), velocity anisotropy, and stellar mass-to-light ratio (\ml), influence the modeled kinematics and the inferred SMBH mass in NGC~4258.  This work represents the fourth dynamical black hole mass measurement obtained with \jwst/NIRSpec, following the initial results for NGC~4736 \citep{Nguyen2025a}, NGC~4486B \citep{Tahmasebzadeh2025}, and UCD~736 \citep{Taylor2025}. Together, these studies demonstrate NIRSpec’s capability to deliver high-quality stellar kinematics and robust dynamical black hole masses.

This paper is organized as follows. In \autoref{sec:obs}, we describe the \jwst/NIRSpec observations, the reduction of the integral field unit (IFU) data, the wiggles correction for a few central spectra, and the extraction of the 2-dimensional (2D) line-of-sight velocity distribution (LOSVD) from the “gold-standard” stellar CO bandhead absorption features. In \autoref{sec:dyn_model}, we combine this 2D LOSVD with the stellar photometric data from \citet{Drehmer2015} as input for the JAM modeling to determine the mass of the central SMBH in NGC~4258. We present and discuss our results, along with concluding remarks, in \autoref{sec:results} and \autoref{sec:conclusion}, respectively.

\section{Observations and Kinematics}\label{sec:obs}

\subsection{JWST NIRSpec IFU}\label{sec:ifu}

The NIRSpec G235H F170LP observation of NGC~4258 was obtained on February 15, 2023, which comprises eight exposures of 218.8 seconds each, yielding a total on-source integration time of 1750.7 seconds. The observations were conducted in high spectral resolution mode ($R\sim2700$), covering the wavelength range of 1.66–3.05 \micron. The NRSIRS2RAPID readout pattern was used, with 14 groups per integration, one integration per exposure, and a four-point medium-cycling dither pattern. Data reduction followed the standard \jwst\ calibration pipeline for NIRSpec IFU data, using STScI pipeline version 1.14.0 and Calibration Reference Data System (CRDS) context 1063\footnote{\url{https://jwst-crds.stsci.edu}}. The final IFU datacube has a spatial pixel scale of 0\farcs1, approximately four times smaller than the SMBH's $r_{\rm SOI}$ if assuming the maser dynamical SMBH mass (\autoref{sec:intro}), and extends within a field-of-view (FoV) of $3\arcsec\times3\arcsec$.

\subsection{Wiggles Corrections} \label{wiggle-correct}

\begin{figure*}
    \centering
    \includegraphics[width=0.93\linewidth]{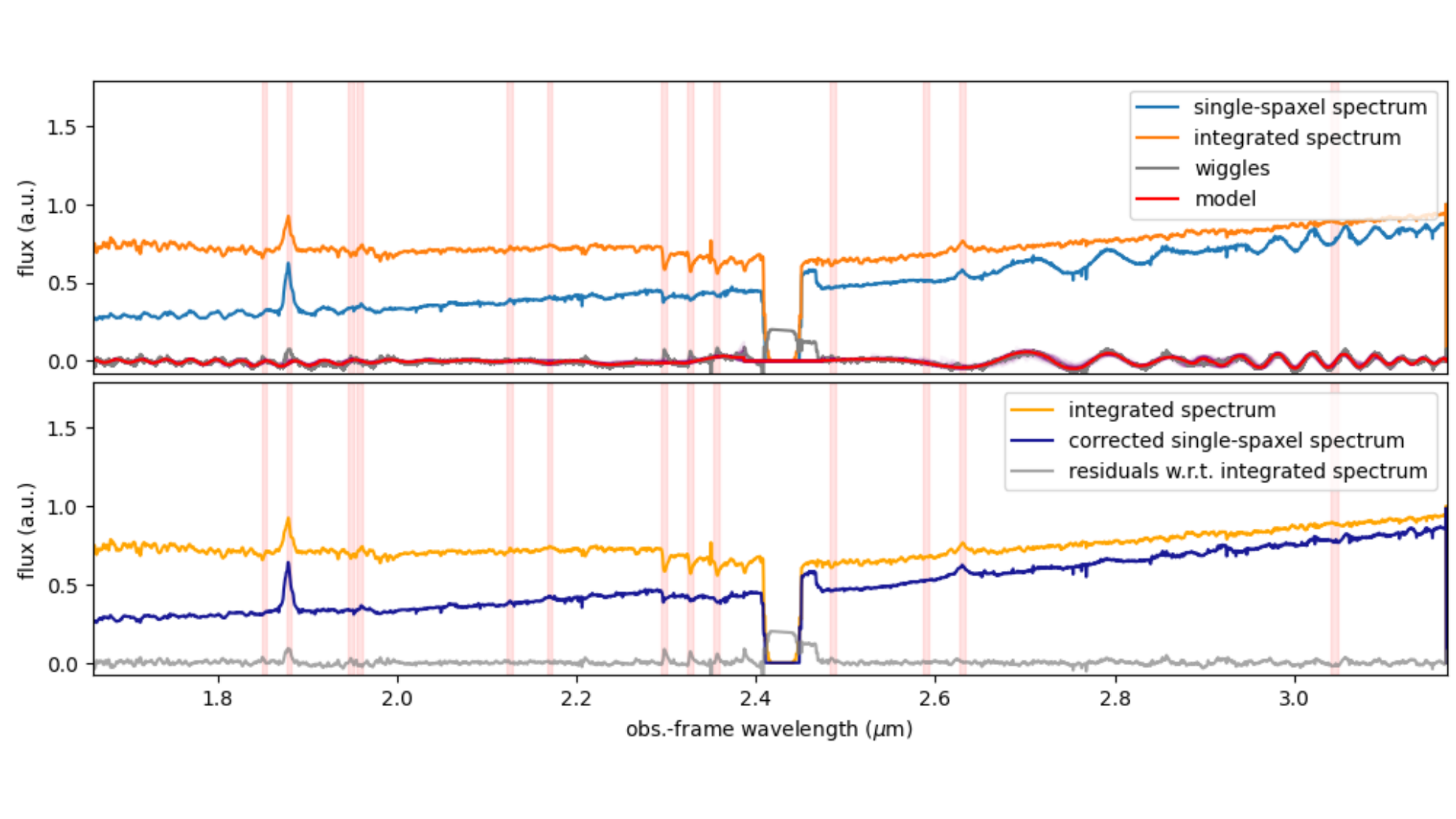}
    \vspace{-8mm}
    \caption{Modeling of the wiggles in single-spaxel spectra for NGC~4258. {\it Top panel:} The integrated spectrum (orange), single-spaxel spectrum (blue), and residual wiggles (gray). The red curve shows the best-fit model to the wiggles. {\it Bottom panel:} The corrected single-spaxel spectrum (dark blue), compared to the integrated spectrum (orange); the gray curve shows the residuals after correction. In all panels, red shaded regions indicate emission lines excluded from the fit.}
    \label{fig:Wiggles}
\end{figure*}

\begin{table}
\centering
\caption{MGE of the NIRSpec G235H/F170LP STPSF PSF}
\begin{tabular}{cccc}
\hline \hline
$j$ & Light fraction & $\sigma (\arcsec)$ & FWHM$_{\rm PSF}^{\rm tot}(\arcsec)$\\
(1) & (2) & (3)    & (4)  \\
\hline\hline
\multicolumn{4}{c}{Extension 1 (fix number of Gaussians = 3)} \\
\hline
1 & 0.6703 & 0.036  &  \multirow{3}{*}{0.08 (PSF 1)} \\
2 & 0.0436 & 0.190   \\
3 & 0.0172 & 0.997   \\
\hline\hline
\multicolumn{4}{c}{Extension 3 (free number of Gaussians)} \\
\hline
1 & 0.7400 & 0.065  & \multirow{5}{*}{0.15 (PSF 2)} \\
2 & 0.1498 & 0.141   \\
3 & 0.0621 & 0.336   \\
4 & 0.0245 & 0.626   \\
5 & 0.0176 & 0.877   \\
\hline\hline
 \multicolumn{4}{c}{Extension 3 (fix number of Gaussians = 3)} \\
\hline
1 & 0.8592 & 0.083  & \multirow{3}{*}{0.20 (PSF 3)} \\
2 & 0.1145 & 0.372   \\
3 & 0.0263 & 1.100   \\
\hline
\end{tabular}
\label{tab:psf} 
\noindent\parbox{\linewidth}{\textbf{Notes.}  Column (1) lists the number of circular Gaussian components. Column (2) shows the fractional light contribution of each Gaussian, normalized to unity. Column (3) gives the Gaussian dispersions along the major axis in arcseconds. Column (4) provides the axial ratios (minor-to-major axis). Column (5) reports the total FWHM of the composite PSF. The FWHM of PSF 2 is consistent with the empirical NIRSpec PSF from \citet{D’Eugenio2024NaAs} and \citet{D'Eugenio2025}.
}
\end{table}

The spatial under-sampling of the NIRSpec PSF produces characteristic fluctuations, commonly referred to as ``wiggles’’, in the single-spaxel spectra of bright targets. The effect is significantly reduced when integrating over apertures larger than $\sim$0\farcs2–0\farcs5 (e.g., \citealt{Law2023}). To mitigate this effect in our data cubes, we apply the correction algorithm developed by \citet{Perna2023},\footnote{\url{https://github.com/micheleperna/JWST-NIRSpec_wiggles}}, which provides an empirical approach to remove wiggles in the absence of a dedicated correction in the standard NIRSpec pipeline. Below we briefly summarise the main steps of the procedure; we refer the reader to \citet{Perna2023} for full technical details.

 \begin{figure*}
 \centering
   	\includegraphics[width=0.95\linewidth]{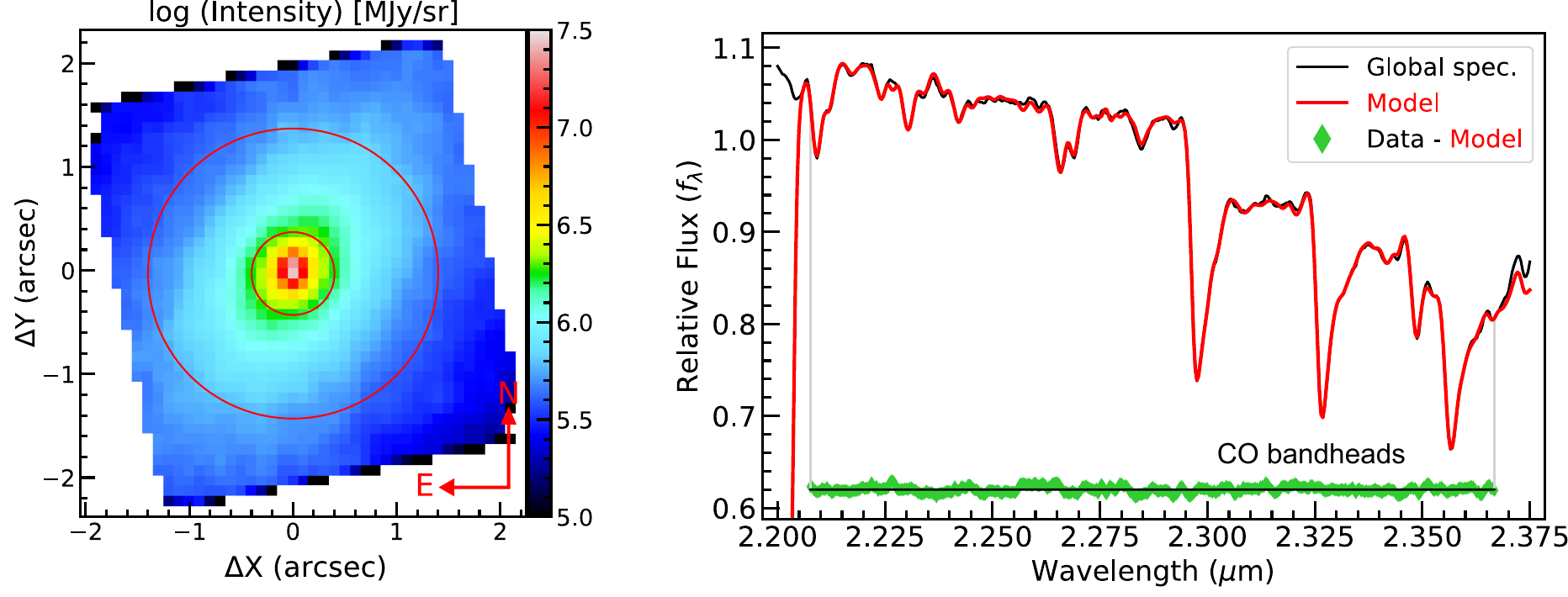} 
	\caption{{\it Left:} Logarithmic integrated intensity map of the NIRSpec G235H/F170LP data cube, collapsed along the spectral axis, excluding the detector gap at $\sim$2.41--2.49~\micron. The annular region outlined by the two red rings ($0\farcs4 < r < 1\farcs4$) indicates the area from which the global spectrum was extracted.     {\it Right:} Optimal stellar template for NGC~4258. The observed global spectrum (black line) is compared to the broadened best-fit template obtained with \textsc{pPXF} (red line). The fit residuals (green points) are vertically offset by +0.62 to compress the $y$-axis range and better illustrate the stellar CO absorption bandheads. This same vertical offset is applied to all subsequent figures of this type.}
	\label{fig:Intensity_maps}
\end{figure*} 

We model the wiggles as a sinusoidal function in the spectrum extracted from the brightest spaxel of the cube, where the wiggles are most prominent across the entire spectral range as shown in \autoref{fig:Wiggles}. Prior to fitting, we mask prominent emission and absorption features, as well as the spectral gap between the two NIRSpec detectors, to avoid contaminating the model with intrinsic spectral features. The modelling of a sinusoidal function is applied iteratively in rolling windows across the spectral range, allowing the model to adapt locally to wavelength-dependent variations. The model is tuned such that the corrected single-spaxel spectrum closely matches the spectrum extracted from a larger aperture, where the wiggles are averaged out. For our targets, we use a circular aperture of 4-pixel radius (0\farcs4), centred on the peak of the collapsed (white-light) image. We verify that this aperture is sufficient to suppress the wiggles in our datasets. The wiggle frequency derived from this reference spectrum is then adopted as a prior for modelling and subtracting the wiggles in other spaxels where the effect is evident.  The IFU data cube corrected for wiggles is used in all subsequent analyses.

\subsection{NIRSpec PSF}\label{sec:nirspec_psf}
 
This work will test the impact of our synthetic NIRSpec PSFs at 2.4~\micron\ on the \Mbh\ measurement in NGC~4258, in addition to the adopted single-Gaussian PSF (FWHM$_{\rm PSF, tot} = 0\farcs15-0\farcs18$) from \citet{D’Eugenio2024NaAs, D'Eugenio2025}.

To generate the synthetic NIRSpec PSFs,  we used {\tt stpsf}\footnote{v2.0.0: \url{https://github.com/spacetelescope/stpsf}} version 2.0.0 (\citealt{Perrin2025}; {\tt stpsf} replaces {\tt WebbPSF}). We used the following parameters: Instrument: NIRSpec, Mode: IFU, Disperser: G235H, Filter: F170LP, Wavelength:  at 2.3~\micron. {\tt stpsf} generates PSFs with native detector sampling and factor 4 oversampling, both with and without the additional instrumental effects (distortion, detector cross-talk, etc.). For our analysis, we used the {\tt DET\_DIST} extension that incorporates the additional instrumental effects, with factor 4 oversampling.

We constructed oversampled synthetic PSF models from extensions 1 and 3 using the Multi-Gaussian Expansion (MGE; \citealt{Emsellem1994, Cappellari2002}) via the \textsc{mge\_fit\_sector} routine in the \textsc{MgeFit} package\footnote{v5.0: \url{https://pypi.org/project/mgefit/}} \citep{Cappellari2002}. For both extensions, we adopted a fixed number of three Gaussians, ensuring each Gaussian fits well with the data at its position. Additionally, we allowed a free number of Gaussians when fitting extension 3, resulting in a total FWHM of $\approx$0\farcs15, fully consistent with the empirical estimate from \citet{D'Eugenio2025}. All three synthetic MGE PSF models are summarized in \autoref{tab:psf}.

\begin{figure*}
    \centering
    \includegraphics[width=0.84\linewidth]{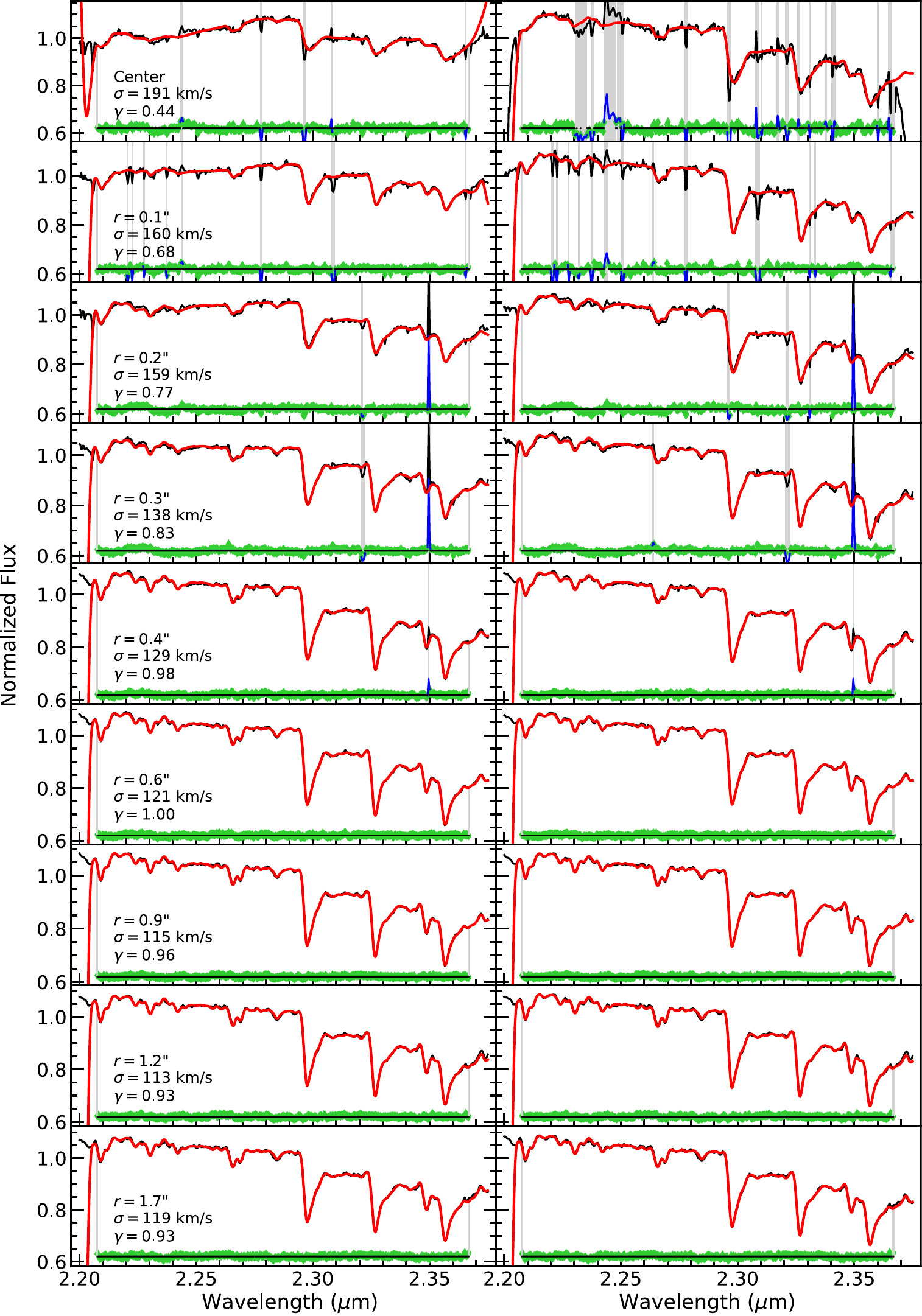}
    \caption{Radial variation in the NIRSpec G235H/F170LP spectrum of NGC~4258. {\it Left column:} Each panel shows the observed spectrum (black line), obtained by coadding spaxels within circular annuli of radius $r$. The best-fitting \textsc{pPXF} model (red line) includes the global stellar template (left panel of \autoref{fig:Intensity_maps}, constrained from \autoref{sec:global_template}) convolved with a Gaussian LOSVD, plus fourth-degree multiplicative and additive polynomials, which represent the nuclear non-thermal spectrum. Residuals are shown as green dots below each fit. Spectral regions affected by emission lines or artifacts were excluded from the fit (gray areas). {\it Right column:} The global stellar template convolved with the LOSVD (red line) is compared to the observed spectrum after subtraction of the nuclear non-thermal component (black line).}
    \label{fig:annuli_spec}
\end{figure*}

\subsection{Stellar Kinematics Templates and pPXF Setup}\label{sec:kine}
 
 Since the nuclear spectra of NGC~4258 are dominated in several central spaxels by AGN continuum emission \citep{Wilkes1995} (after applying wiggles correction; \autoref{wiggle-correct}), care must be taken in the stellar kinematics extraction. The AGN continuum itself does not bias the kinematics, but it dilutes the stellar CO bandhead absorption features of cool, evolved giant stars that are used to derive the stellar velocity dispersion $\sigma$. This dilution reduces the measured line strength $\gamma$ and, if uncorrected, could lead to underestimates of $\sigma$ \citep[see Section~2.2 of][for the strong correlation between $\sigma$ and $\gamma$]{vanderMarel1993}. In addition, the rising AGN continuum could be mistaken for a stellar population change, distorting the template mix and introducing systematic errors. To avoid these issues, it is essential to include an additive polynomial in the spectral fits to account for the smooth AGN continuum component while preserving the stellar absorption features \citep{Cappellari2009}.  

For clarity, the observed line-strength $\gamma$ can be defined as the stellar flux fraction,
$\gamma ={\rm stars}/({\rm stars + AGN})$,
i.e., the ratio between the flux contributed by the best-fitting stellar template alone and the observed mean flux in the fitted spectral range. Intuitively, $\gamma \approx 1$ in regions far from the nucleus where the AGN contribution is negligible, and it decreases below unity toward the center, dropping substantially in the few spaxels where the AGN continuum dominates.

In this work, we extracted 2D stellar kinematic maps from the NIRSpec G235H/F170LP datacube of NGC~4258 using the CO bandheads absorption features \citep[e.g.,][]{Cappellari2009, Nguyen2014, Nguyen2023, Thater2023} and the \textsc{Penalized PiXel-Fitting} (\textsc{pPXF}) method\footnote{v8.2.1: \url{https://pypi.org/project/ppxf/}} \citep{Cappellari2004,Cappellari2017, Cappellari2023}. To evaluate the robustness of our results and quantify template-related systematics, we employed both the empirical X-shooter Spectral Library Data Release 3 (XSL DR3; 830 spectra from 683 stars\footnote{\url{http://xsl.u-strasbg.fr}}; \citealt{Verro2022}) and the PHOENIX synthetic stellar library\footnote{\url{http://phoenix.astro.physik.uni-goettingen.de}} \citep{Husser2013}.

The XSL DR3 library covers the full wavelength range of the X-shooter spectrograph (3000–25,000~\AA) with a spectral resolution of $R \approx 10{,}000$, and has been corrected for dust extinction. It includes a wide variety of stellar types, from O to M, including asymptotic giant branch stars, and spans most of the Hertzsprung–Russell diagram. In contrast, the PHOENIX library, generated using the spherical mode of PHOENIX, which includes the effects of microturbulence in stellar atmospheres, offers significantly higher spectral resolution ($R \approx 500{,}000$ in the optical and near IR, $R \approx 100{,}000$ in the IR and $\Delta\lambda=0.1$ \AA), extends from 0.3 to 5~$\mu$m, and is free from telluric absorption gaps. It spans a broad range of stellar parameters: $2{,}300$~K$\leq T_{\rm eff} \leq 25{,}000$~K, $0.0 \leq \log(g) \leq +6.0$, $-4.0 \leq [{\rm Fe/H}] \leq +1.0$, and $-0.2 \leq [\alpha/{\rm Fe}] \leq +1.2$. 

The XSL will be adopted throughout the main text as our fiducial template unless otherwise noted.

\begin{figure}
    \centering
    \includegraphics[width=0.95\linewidth]{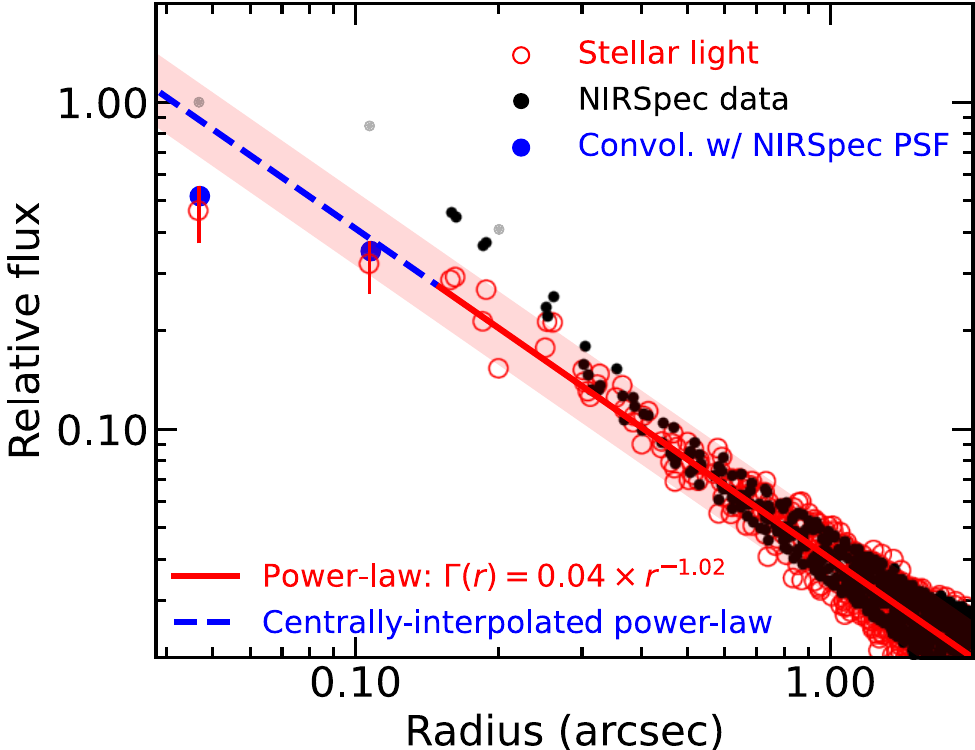}
    \caption{The radial surface brightness profile $I(r)$, measured from individual \textsc{Voronoi} bins in the NIRSpec G235H/F170LP datacube of NGC~4258 (filled black dots), is compared to the estimated stellar light profile $\Gamma(r) = I(r)\gamma(r)$ (red open circles). The underlying stellar distribution is smooth and well approximated by a single-power-law (red line with pink region shows its $1\sigma$ uncertainty). The PSF–convolved central interpolation of this single power-law also reproduces the two innermost $\gamma$ measurements, which lie below the intrinsic (unconvolved) profile, indicating that their apparent decline is fully consistent with PSF effects. The gray points represent the pixels where $\gamma < 0.5$.} 
    \label{fig:stellar_perc}
\end{figure}

\begin{table}[]
\centering
\caption{JWST/NIRSpec Kinematic Data of the NGC 4258 Nucleus}
\vspace{-4mm}
\begin{tabular}{cccccccc}
\hline \hline
$\Delta$R.A. & $\Delta$Decl. & $V$ & $\Delta V$ & $\sigma$ & $ \Delta\sigma$ \\
($\arcsec$) & ($\arcsec$) & (\kms) & (\kms) & (\kms) & (\kms) & \\
\hline
$-$0.168 & 0.130 & 39.04 & 2.45 & 132.75 & 3.01 \\ 
$-$0.085 & 0.185 & 18.58 & 2.48 & 139.92 & 3.07 \\ 
$-$0.030 & 0.101 & 21.50 & 5.28 & 171.38 & 6.69 \\ 
$-$0.114 & 0.047 & 36.20  & 3.56 & 156.29 & 4.47 \\ 
     0.054 & 0.156 & $-$6.27 & 5.46 & 139.51 & 6.75 \\ 
$-$0.001 & 0.240 & 4.54 & 2.64 & 136.65 & 3.27 \\ 
     0.025 & 0.018 & $-$5.36 & 8.07 & 191.06 & 9.52 \\ 
...            & ...       & ... & ... & ... & ... \\ \hline
\end{tabular}
\noindent\parbox{\linewidth}{This table is available in its entirety in machine-readable form.  A portion is shown here for guidance regarding its form and content. A data copy is also available in Zenodo at \dataset[doi: 10.5281/zenodo.17729122]{https://doi.org/10.5281/zenodo.17729122}.}
\label{tab:kin}
\end{table}

\subsection{Determining the Optimal Stellar Template}\label{sec:global_template}

To separate the stellar light from the non-thermal AGN continuum in the NIRSpec G235H/F170LP spectra of NGC~4258, we need a fixed optimal XSL template for all spatial positions to ensure an unbiased extraction of stellar kinematics \citep{Marconi2000, Silge2003}. Since there is no clear evidence of nuclear stellar population variation \citep{Siopis2009}, we assumed the fixed optimal XSL template and modeled the non-thermal AGN continuum using both multiplicative and additive Legendre polynomials \citep[e.g.,][]{Simon2024} within \textsc{pPXF}.  We carefully experimented with different combinations of the degrees of the multiplicative and additive Legendre polynomials and adopted as our preferred approach a fit with {\tt mdegree = 4} and {\tt degree = 4}, allowing for variations in stellar line strength with radius and accounting for residual sky subtraction errors, spectral calibration imperfections, and AGN contamination.  The polynomials account for low-order variations in line strength and potential instrumental effects, allowing velocity dispersion to be reliably measured a few central spaxels (e.g., $0\farcs1-0\farcs3$), before AGN photon noise overwhelms the stellar signal \citep[e.g.,][]{Cappellari2009}.  

To obtain a global spectrum, we combined all spaxel spectra within an annular region from 0\farcs4 to 1\farcs4 centered on the centrally brightest pixel (outlined by the two red rings in the left panel of \autoref{fig:Intensity_maps}), excluding the central spaxels that are strongly contaminated by AGN continuum emission. This annular spectrum achieves a signal-to-noise ratio (S/N) of 215 per spectral pixel. Prominent emission lines were masked, and the spectrum was logarithmically rebinned along the spectral dimension using a constant velocity scale by setting {\tt velscale = 50} \kms\ per pixel, as calculated using Eq. 8 of \citet{Cappellari2017}. Next, we took into account the instrumental broadening in the XSL template by convolving the XSL template with a Gaussian whose dispersion is defined as the differential Gaussian dispersion between the template spectra and the global spectrum \citep[Eq. 5 of][]{Cappellari2017}. Here, when deriving the fixed optimal XSL template, we adopt a constant Gaussian. However, when deriving the kinematic bins across the NIRSpec FoV (\autoref{sec:kinematics}),  where the line broadening may vary spatially and spectrally, we convolve the XSL template with a spatial and wavelength-dependent line spread function \citep[e.g.,][]{Nguyen2018} derived by \citet[][Appendix B]{Nguyen2025a}.

\begin{figure*}
    \centering
    \includegraphics[width=0.98\textwidth]{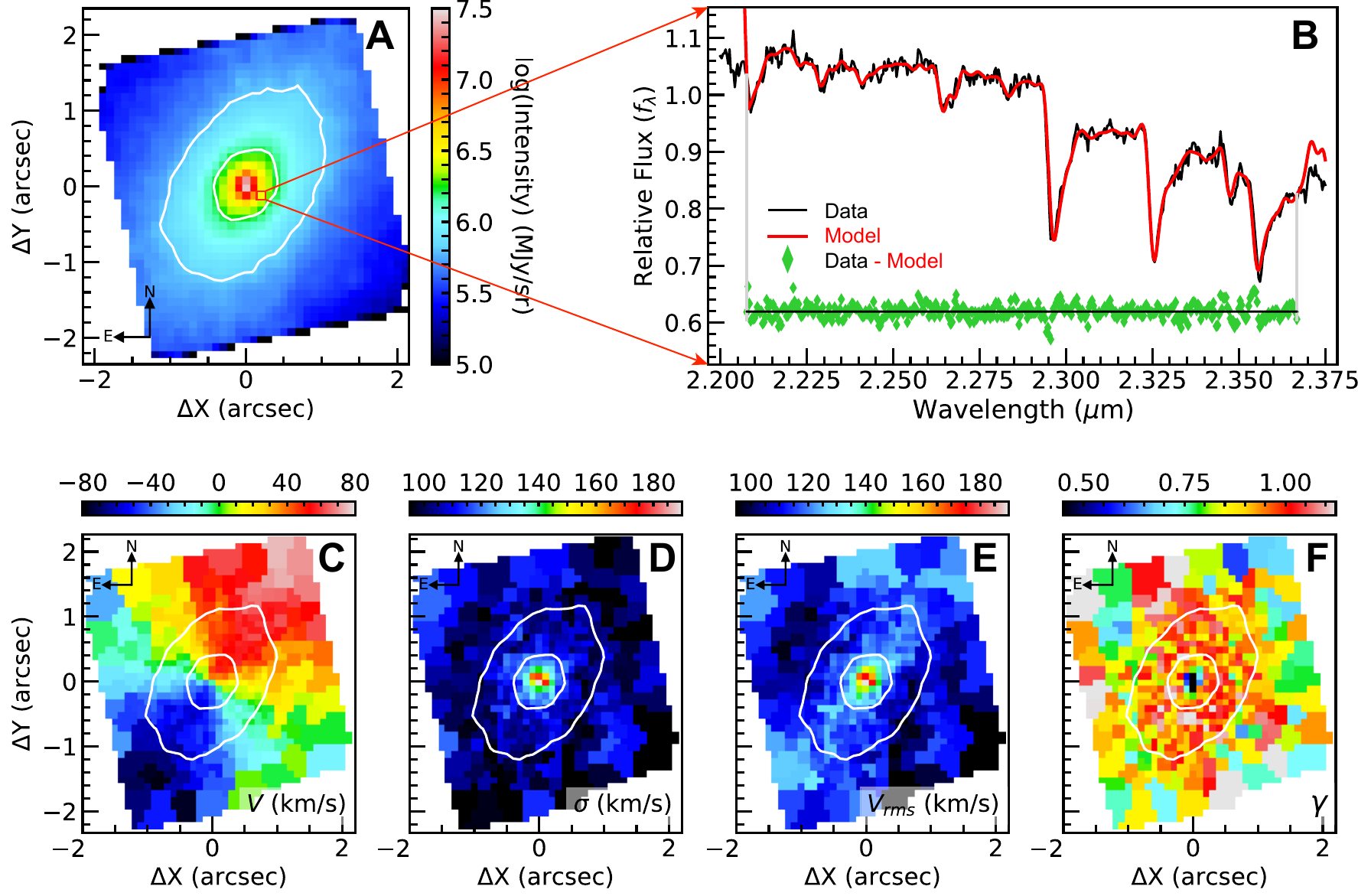}
    \caption{Stellar kinematics extracted from the \jwst/NIRSpec G235H F170LP IFU of NGC~4258 are presented. {\it Panel A:} Same as the left panel of \autoref{fig:Intensity_maps}.  {\it Panel B} displays a representative \textsc{pPXF} fit for a central-offset \textsc{Voronoi} bin that is unaffected by wiggle artifacts; its location is marked in Panel A. The observed spectrum, which includes the stellar CO band heads absorption features at $\sim$2.3 \micron, is shown in black, with the best-fit empirical XSL template overlaid in red. The fit residuals, {\tt (data - model)}, are shown in green. The vertical gray lines indicate the spectral range used in fitting the templates to the spectra in all bins. {\it Panels C–F:} present maps of the stellar rotation velocity ($V$), velocity dispersion ($\sigma$), root-mean-square velocity ($V_{\rm rms} = \sqrt{V^2 + \sigma^2}$), and stellar light contribution fraction ($\gamma$). White contours trace the intensity distribution, decreasing by 1 magnitude per arcseconds$^{2}$ from the center outward.} 
    \label{fig:kin-map}
\end{figure*}

We then applied \textsc{pPXF} to fit this global spectrum using the XSL-instrumental-broadened spectra and a Gauss–Hermite expansion truncated at second order by setting {\tt moments = 2}. This setup yields robust estimates of the recession velocity ($V$, or rotational velocity relative to the systemic velocity $V_{\rm sys}$) and velocity dispersion ($\sigma$), which are our primary kinematic quantities of interest. Modeling only $V$ and $\sigma$ avoids the need to select a bias parameter for penalizing higher-order Gauss–Hermite moments \citep{Gerhard1993, vanderMarel1993}, which otherwise requires tuning via Monte Carlo simulations.

This fitting process produced a fixed optimal stellar template, defined as a linear combination of 18 giant stars' spectra. The template provides an accurate description of the spectrum from 2.200 to 2.375 \micron, which contains most of the significant stellar absorption features of the CO bandheads, as seen in the right panel of \autoref{fig:Intensity_maps}.  The template is dominated by the K4III star (HD109871 of XSL), which contributes $\approx$50\% of the total flux and alone provides a relatively good fit to the annular spectrum. However, additional contributions from a giant M9 star (BMB 289, $\approx$12\% of the flux) and an M4 star (SHV 0535237-700720, $\approx$8\% of the flux) are required to reproduce the data. This best-fit \textsc{pPXF} model for the integrated spectrum within an annular region of $0\farcs4 < r < 1\farcs4$ yields a stellar velocity dispersion of $\sigma = 104 \pm 4$~\kms\ and a Heliocentric recession velocity of $V = 451 \pm 5$~\kms, consistent with the previously reported value for the bulge component constrained from Gemini/NIFS observations \citep{Drehmer2015}.  These 1$\sigma$ measurement errors were determined as the standard deviation through 200 Monte Carlo perturbation simulations of the best-fit model \citep{Hoaglin1983, Cappellari2004}. 

\subsection{Separating Stellar Light from AGN Continuum}\label{sec:non-thermal}

To assess the reliability of the fixed global stellar template for extracting stellar kinematics while accounting for the non-thermal continuum, we fit high-S/N spectra constructed by coadding spaxels within concentric annuli spanning $0\farcs1$–$1\farcs7$ in the NIRSpec G235H/F170LP datacube using \textsc{pPXF}. 

In \autoref{fig:annuli_spec}, we present the radial variation of the NIRSpec G235H/F170LP spectra for NGC 4258 and their corresponding best-fitting \textsc{pPXF} models. The measurements in the left-hand panels show the change in stellar line-strength ($\gamma$) and velocity dispersion ($\sigma$) with radius, while the right-hand panels display the fractional flux contribution of the best-fitting stellar template over the same radial range.  The influence of the non-stellar continuum is evident from the varying slope of the spectra. At the nucleus, the non-thermal component rises modestly, producing a nearly linear continuum in the central spaxel, where it contributes 56\% of the total flux. This fraction drops rapidly to $\approx$32\% at $r \approx 0\farcs1$ and falls below 23\% beyond $r = 0\farcs2$. Despite this non-stellar contribution, the stellar spectrum at all radii is still well represented by the fixed, convolved stellar template after subtracting the continuum.

We quantified the dilution of stellar light by the central non-thermal continuum in \autoref{fig:stellar_perc}, which shows the surface brightness profile $I(r)$ measured from the NIRSpec G235H/F170LP datacube of NGC~4258, along with the radial profile of the stellar contribution, estimated as $\Gamma(r) = I(r)\gamma(r)$ based on the \textsc{pPXF} fits in individual spatial bins (see Figure 4 of \citealt{vanderMarel1994} for M87 and Figure 6 of \citealt{Cappellari2009} for Centaurus~A).  We approximated the radial stellar-light distribution with a single power-law function, $\Gamma(r) = \alpha\times r^\beta$ (red line), adopting $\alpha = 0.04$ and $\beta = -1.02$. This single power-law overpredicts the stellar-light fraction within $0\farcs1$ if the effect of the \jwst/NIRSpec PSF at 2.3~$\mu$m is not included. To properly account for the PSF effect, we performed the following steps:
\begin{itemize}
\item We generated a synthetic image with an intrinsic surface-brightness profile $\Sigma \propto r^{-1.02}$ sampled at 0\farcs025 pixel$^{-1}$, i.e., oversampled by a factor of four relative to the native \jwst/NIRSpec scale (\autoref{sec:nirspec_psf}).
\item We convolved this image with the synthetic \jwst/NIRSpec PSF at the same sampling.
\item We rebinned the convolved image to the NIRSpec scale (0\farcs1 pixel$^{-1}$) by summing $4 \times 4$ spaxels.
\item We measured the stellar fraction in the central spaxel of the rebinned image.
\end{itemize}
The resulting central-stellar-fraction measurement falls along the same trend defined by the two innermost $\gamma$ values as seen in \autoref{fig:stellar_perc}, showing that the apparent break is entirely explained by PSF convolution. Therefore, the deviation near the nucleus does not indicate a true change in the intrinsic stellar surface-brightness profile within the NIRSpec FoV, but instead reflects instrumental and sampling effects that are mitigated once the PSF is properly accounted for.

This experiment confirms that the observed radial spectral variation is consistent with weak non-thermal continuum emission and residual instrumental artifacts, rather than intrinsic changes in the stellar population. The measurement of $\gamma(r)$ enables a reliable assessment of the stellar kinematics even in the unresolved nucleus, where the stellar light is strongly diluted by AGN-related emission.

Given that stars still contribute 44\% of the total flux in the central spaxel, the kinematics there are subject to larger systematic uncertainties ($\approx$10\%) but remain sufficiently reliable to be included in our dynamical modeling (\autoref{sec:jam}). Beyond $r = 0\farcs1$, the stellar kinematics are robust and can be measured with uncertainties below 4\%.

\subsection{Two-Dimensional Kinematic Maps}\label{sec:kinematics}

To derive the LOSVD maps of the nucleus of NGC~4258, we applied the adaptive \textsc{Voronoi} binning method\footnote{v3.1.5: \url{https://pypi.org/project/vorbin/}} \citep{Cappellari2003} to spatially group spaxels until reaching a target-S/N = 100 per spectral pixel, resulting in $N=403$ kinematic \textsc{Voronoi} bins. The chosen target S/N represents a compromise: it allows spaxels beyond $r \approx 0\farcs5 \approx r_{\rm SOI}$ to be grouped, while retaining unbinned spaxels within this radius to enable precise measurements of the stellar kinematics inside $r_{\rm SOI}$.  Inside the SOI, the S/N of individual unbinned spaxels ranges from 100 to 200 per spectral pixel, increasing toward the central peak.   Each binned/unbinned spectrum was resampled onto a logarithmic wavelength scale. For all bins, the \textsc{pPXF} fit used the fixed optimal global stellar template, convolved with the LOSVD, and included fourth-degree multiplicative and additive Legendre polynomials.

\autoref{fig:kin-map} presents the logarithmic integrated intensity map (panel A, same as the left panel of \autoref{fig:Intensity_maps}), with a marked spaxel that has the spectral fit \textsc{pPXF} in panel B, as well as the resulting maps of rotational velocity $V$ (panel C), velocity dispersion $\sigma$ (panel D), root-mean-square velocity $V_{\rm rms} = \sqrt{V^2 + \sigma^2}$ (panel E), and stellar line-strength $\gamma$ (panel F).   The nucleus exhibits modest rotation, with $\vert V \vert  \approx 80\pm3$ \kms\ at the edge of the NIRSpec IFU field (after subtracting the systemic/Heliocentric velocity of $v_{\rm sys} = 451 \pm5 $ \kms), and a global kinematic position angle of PA$_{\rm kin} = 147^\circ \pm 7^\circ$. Both $v_{\rm sys}$ and PA$_{\rm kin}$ are derived from the velocity field using the \textsc{pafit}\footnote{v2.0.8: \url{https://pypi.org/project/pafit/}} package \citep{Krajnovic2006}. The velocity dispersion rises gradually from $\sim$$105\pm5$ \kms\ at $r \gtrsim 1\arcsec$ to a sharp central peak of $\sim$$191\pm 12$ \kms\ within $r \lesssim 0\farcs2$, consistent with a corresponding increase in $V_{\rm rms}$ from $\sim$$120\pm4$ \kms\ to $\sim$$191\pm12$ \kms\ along the nucleus' major axis. This steep rise in velocity dispersion (and root-mean-square velocity) strongly supports the presence of a central SMBH. We presented these stellar kinematic measurements in \autoref{tab:kin}.

We also assessed the robustness of our stellar kinematic measurements by reanalyzing the data using the synthetic PHOENIX stellar library (\autoref{sec:kine}) and compared the results to those obtained with the empirical XSL library (panels C–F of \autoref{fig:kin-map}). The resulting maps are highly consistent, with differences $\lesssim$3\%.

The irregular morphology of the stellar-light fraction in panel F of \autoref{fig:kin-map} is not physical but is caused by residual wiggle/continuum effects and the sensitivity of $\gamma$ to the adopted \textsc{Voronoi} binning. These small residuals introduce bin-to-bin variations in the continuum subtraction, producing the apparent patchiness and the decrease/scatter of $\gamma$ in some outer bins despite their adequate S/N. Importantly, this behavior does not affect our stellar kinematic measurements. In the \textsc{pPXF} fits, the additive and multiplicative polynomials effectively absorb the AGN continuum and any remaining wiggle/continuum structure, preventing the known $\sigma$–$\gamma$ dependency, between the velocity dispersion and the line strength of the stellar templates, and ensuring that the derived velocity dispersions are robust across the field.

Our derivation of the stellar LOSVD from the NIRSpec G235H/F170LP IFU data is consistent with the Gemini/NIFS measurements of \citet{Drehmer2015}. The central $V_{\rm rms}$ from NIRSpec is $\approx$11 \kms\ higher than the NIFS value; this small difference is unsurprisingly given the substantially different PSFs of JWST/NIRSpec and Gemini/NIFS, which change the amount of central velocity smoothing and can produce modest systematic differences in $V_{\rm rms}$.

\subsection{Defining the Galaxy Center}\label{sec:photo_cent}

We identified the photometric AGN-aligned center by computing the weighted barycenter of the $\sim$100 brightest pixels. Unlike the kinematic center, which is less accurately determined because it relies on LOSVD extractions (\autoref{sec:kinematics}), the photometric center can be measured with much higher precision. Given that the black hole is expected to coincide with the surface-brightness center, we therefore adopted the photometric AGN-aligned center to define the origin of our models, and this choice is reflected in all figures.

\section{Dynamical Modelling}\label{sec:dyn_model}

\subsection{Stellar Mass Model}\label{sec:sb}
 
The surface brightness of NGC~4258 was carefully modeled by \citet{Drehmer2015} using $K$s-band photometry from the Two Micron All Sky Survey (2MASS; \citealt{Jarrett2003}) to constrain the extended stellar light beyond the central $19\farcs2 \times 19\farcs2$ region. Within this region, the stellar light distribution was derived from the \hst/Near Infrared Camera and Multi-Object Spectrometer (NICMOS) F222M image (PID: 7230, PI: Scoville). The resulting surface brightness model in terms of an MGE \citep{Emsellem1994, Cappellari2002}, which consists of 12 concentric Gaussians, listed in Table 1 of \citet{Drehmer2015}, based on the \textsc{mge\_fit\_sector} routine \citep{Cappellari2002}.

To convert the $K$s-band surface brightness distribution into a stellar mass density model, we adopted the mass-follows-light assumption by scaling the MGE with a dynamical \ml\ in the $K$s band (\ml$_K$). The dark matter fraction is expected to be completely negligible in the dense nuclear region of this Galaxy. The MGE was analytically deprojected assuming a free inclination angle ($i$) to yield a 3-dimensional (3D) axisymmetric mass density distribution. To account for the gravitational potential of the central SMBH, we approximated its mass distribution as a proper Keplerian potential by setting \texttt{analytic\_los=False}.  This is justified by the relatively large SOI of the SMBH in NGC~4258 (\autoref{sec:smbh}), which extends to a radius $\approx$4–5 times greater than the NIRSpec pixel scale and $\approx$2.5 times larger than the NIRSpec FWHM$_{\rm PSF}$.

\subsection{Jeans Anisotropic Models}\label{sec:jam}

\begin{figure*}[p]
\centering
\includegraphics[width=0.98\textwidth]{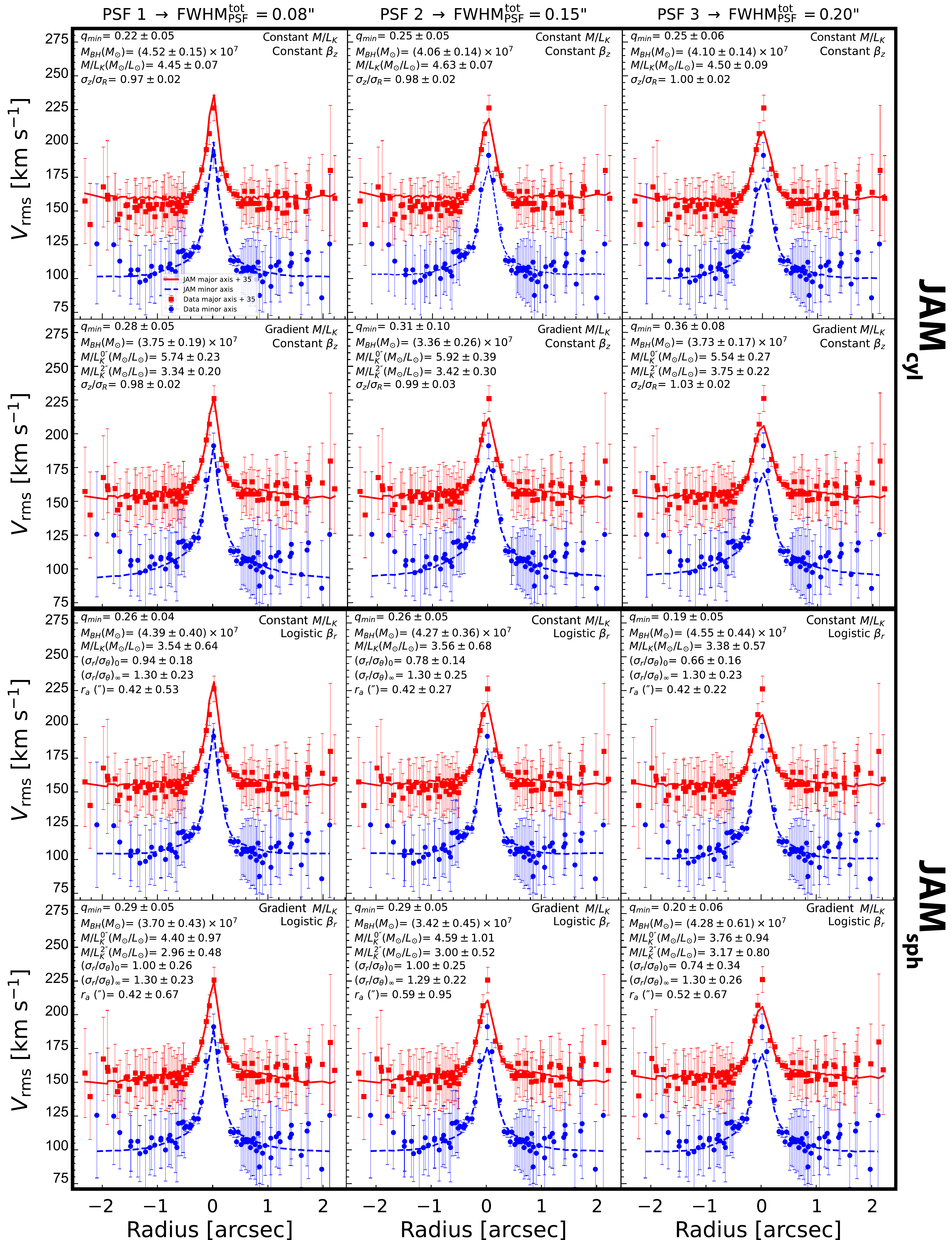}
\caption{Stellar kinematics extracted from the NIRSpec G235H/F170LP data cube are shown as filled red (major axis + 35 \kms) and blue (minor axis) points, overlaid with best-fit JAM models assuming various combinations of \ml$_K$ and orbital anisotropy. Each model includes its corresponding best-fit parameters and associated 1$\sigma$ uncertainties in the legend.}
\label{fig:1D-vrms-cyl}
\end{figure*}               

In this work, we investigate the impact of the axisymmetric gravitational potential under two extreme assumptions for the velocity ellipsoid alignment on the dynamical determination of the central \Mbh\ in NGC~4258, using new stellar kinematic measurements from the NIRSpec G235H/F170LP data. We compare the Jeans Anisotropic Model with cylindrically aligned velocity ellipsoids \citep[\jamcyl;][]{Cappellari2008} and the version with spherically aligned velocity ellipsoids \citep[\jamsph;][]{Cappellari2020}.

The \jamcyl\ model assumes a velocity ellipsoid aligned with cylindrical coordinates ($R, \phi, z$), allowing for anisotropy in the `vertical' $z$ direction, in which the vertical velocity dispersion is different from the equal radial and tangential dispersions ($\sigma_z \ne \sigma_R = \sigma_\phi$, velocity ellipsoid axially symmetric around the vertical direction: $\beta_z(R,z)=1-(\sigma_z/\sigma_R)^2$). In contrast, the \jamsph\ model assumes a velocity ellipsoid aligned with spherical coordinates ($r, \theta, \phi$), allowing for anisotropy in the radial direction, in which the radial velocity dispersion is different from the equal azimuthal and tangential velocity dispersions ($\sigma_r \ne \sigma_\theta = \sigma_\phi$, or velocity ellipsoid axially symmetric around the radial direction: $\beta_r(r,\theta)=1-(\sigma_\theta/\sigma_r)^2$).

These \jamcyl\ and \jamsph\ models are achieved by setting the keyword {\tt align=`cyl'} to enforce cylindrical alignment and {\tt align=`sph'} to enforce spherical alignment in the \texttt{jam.axi.proj} procedure within the \textsc{JamPy}\footnote{v7.2.0: \url{https://pypi.org/project/jampy/}} package \citep{Cappellari2020}, approximating for the intrinsic second velocity moments (i.e., $\langle\sigma_z^2\rangle$ and $\langle\sigma_\phi^2\rangle$ for \jamcyl; or $\langle\sigma_r^2\rangle$ and $\langle\sigma_\theta^2\rangle$ for \jamsph). This approximation is purely a numerical convenience that allows for fast computation of the model predictions for \jamcyl, while the \jamsph\ models always compute the BH potential exactly.

\subsection{Model Grid and Parameter Space}\label{sec:12jam}

To model the stellar kinematics and dynamics of the NGC~4258's nucleus, we compared the observed NIRSpec G235H/F170LP $V_{\rm rms}$ map with predictions from both \jamcyl\ and \jamsph\ models.  These models estimate \Mbh\ (sampled logarithmically) and constrain additional parameters scaled linearly: the mass-to-light ratio (\ml$_K$), inclination ($i$; converted to $q_{\rm min}$), and orbital anisotropy. For \jamsph, we assumed radial anisotropy ($\beta_r$), parametrized as the ratio $\sigma_\theta/\sigma_r$; for \jamcyl\, we assumed vertical anisotropy ($\beta_z$), parametrized as $\sigma_z/\sigma_R$. In the \jamcyl\ model, we adopted a constant anisotropy without priors. For the physically motivated \jamsph\ model, we implemented the \citet{Simon2024} logistic anisotropy with specific priors (\autoref{sec:logistic_anisotropy}). These two limiting cases assume a velocity ellipsoid aligned either radially (\jamsph) or vertically (\jamcyl), which shapes the predicted LOSVD. 

In addition to the two JAM models, we assessed the systematic uncertainties in the \Mbh\ measurement of NGC~4258 using three synthetic JWST/NIRSpec G235H/F170LP PSFs (\autoref{tab:psf}) and by testing both constant and varying \ml$_K$ models (\autoref{sec:default_mlvary}). In total, we performed 12 JAM model runs as listed in \autoref{tab:adamet}, which together bracket the main sources of systematic uncertainty in our dynamical analysis, and summarized below:
\begin{enumerate}
    \item \textbf{Anisotropy alignment (2 options):}  
    Either cylindrically aligned velocity ellipsoid (\jamcyl) or spherically aligned one (\jamsph).
    \item \textbf{Point Spread Function (3 options):}  
    Three {\tt stpsf} synthetic JWST/NIRSpec PSFs were tested (\autoref{tab:psf}).  
    \item \textbf{Mass-to-light ratio (2 options):}  
    Either a constant \ml$_K$ or a radially varying \ml$_K$ (\autoref{sec:default_mlvary}).
\end{enumerate}

\subsection{Linearly Varying M/L$_K$ Profile}\label{sec:default_mlvary}

\citet{Siopis2009} demonstrated that the color gradient in the outer disk of NGC~4258 ($r > 2\arcsec$) is negligible. As shown in their Figure 4, there is no significant difference among the surface brightness profiles in the $V$, $R$, and $K$ bands extracted along the galaxy’s photometric major axis after correcting for dust lanes and AGN contamination. This suggests that $M/L_K$ remains constant beyond $2\arcsec$, i.e., $M/L_K(r \ge 2\arcsec) = M/L_K^{2\arcsec}$. However, within this radius, the presence of non-thermal AGN emission likely causes a variation in $M/L_K$ profile toward the galaxy center, which we model by allowing for a different nuclear mass-to-light ratio $M/L_K^{0\arcsec}$. We thus performed alternative JAM tests assuming a radially varying $M/L_K(r)$, which changes linearly from $r = 0\arcsec$ to $r = 2\arcsec$ and remains constant outside this interval.

In JAMs, the $M/L_K(r)$ profile is implemented by associating a different $(M/L_K)_j$ to each Gaussian component with dispersion $\sigma_j$ in the MGE listed in Table 1 of \citet{Drehmer2015}, as follows: 
\begin{equation}
(M/L_K)_j = 
\begin{cases}
    M/L_K^{0\arcsec} + \dfrac{M/L_K^{2\arcsec} - M/L_K^{0\arcsec}}{2\arcsec} \times \sigma_j, & \sigma_j < 2\arcsec\\
    M/L_K^{2\arcsec}, & \sigma_j \ge 2\arcsec
\end{cases}
\end{equation}

\begin{figure*}
\centering
\includegraphics[width=0.8\textwidth]{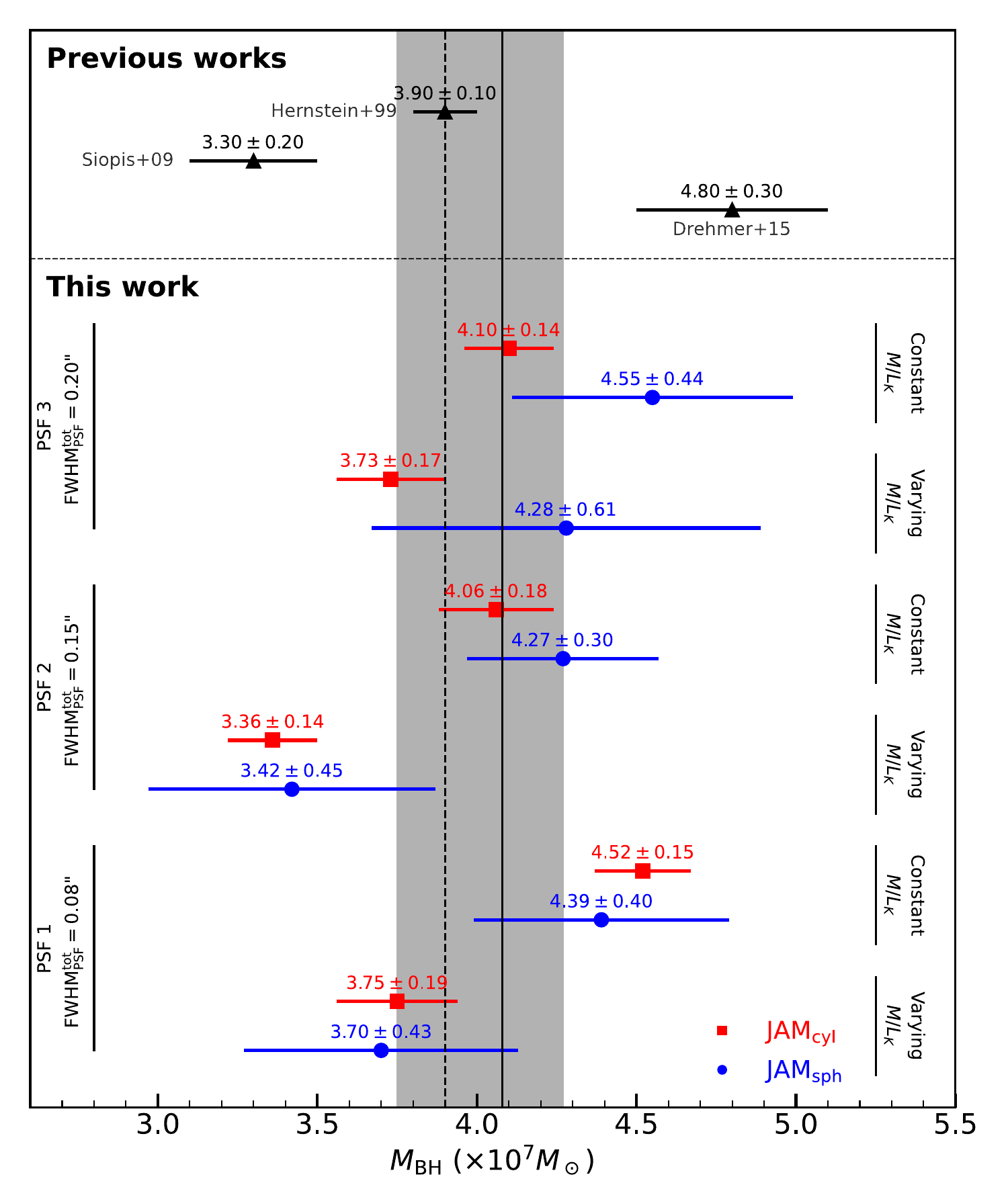}
\caption{Black hole mass measurements and their 1$\sigma$ uncertainties for NGC 4258 are shown from our 12 JAM model runs based on the NIRSpec G235H/F170LP data cube, compared with previous dynamical estimates from ``golden standard'' masers \citep[vertical dashed line;][]{Herrnstein1999}, stars \citep{Siopis2009, Drehmer2015}, and gas \citep{Pastorini2007}. The legend indicates the results from our JAM models under different assumptions for the NIRSpec PSF, stellar velocity anisotropy, and $M/L_K$ profiles. Our adopted \Mbh\ is shown as the vertical solid line, with the gray band indicating the ensemble median and 68\% (1$\sigma$) bootstrap confidence interval from all 12 JAM models.}
\label{fig:summary_all_BH_ngc4258}
\end{figure*}

\subsection{Logistic Anisotropy for \jamsph\ Models}\label{sec:logistic_anisotropy}                

We further explored the \jamsph\ model using a radially varying anisotropy \citep{Simon2024}, obtained using the \texttt{logistic} keyword in \jamsph, to investigate how the stellar orbital structure changes across the nuclear region of NGC~4258 by adopting the logistic function of $\log r$ below:
\begin{equation}
\beta_r(r) = \beta_{r, 0} + \frac{\beta_{r,\infty} - \beta_{r,0}}{1 + (r_a/r)^\alpha}
\end{equation}
Here, $r_a$ is the transition radius at which the stellar motion changes from being dominated by $\beta_{r, 0}$ to $\beta_{r,\infty}$, which represent the anisotropy parameters at the center and at large radii, respectively. These parameters relate to the ratio of stellar velocity dispersions in the radial and azimuthal directions at the corresponding locations as:
\begin{equation*}
(\sigma_r/\sigma_\theta)_0 = \frac{1}{\sqrt{1 - \beta_{r,0}}} \;\; {\rm and} \;\;
(\sigma_r/\sigma_\theta)_\infty = \frac{1}{\sqrt{1 - \beta_{r,\infty}}}
\end{equation*}
We applied specific priors $0.5<(\sigma_r/\sigma_\theta)_0<1$ and $1<(\sigma_r/\sigma_\theta)_\infty<1.3$ to constrain the inner and outer anisotropies, as suggested in \citet[table~3]{Simon2024}. Moreover, the anisotropy transition radius $r_a$ was restricted to lie between the range of $r_{\rm SOI} < r_a < $1\arcsec\ to prevent the model from becoming radially anisotropic within $r_{\rm SOI}$. Additionally, we fixed the sharpness parameter to $\alpha=2$ to reduce the dimension of parameter space and reduce degeneracies, given the weak dependence of the results on the choice of this parameter.

\subsection{MCMC Fitting and Uncertainty Estimation}\label{sec:smbh}

\begin{table*}
\centering
\caption{Summary of the 12 JAM models best-fitting parameters and \emph{formal} uncertainties.}
\vspace{-2mm}
\begin{tabular}{l c c c c c c c c c}
\hline\hline
$M/L_K$ & FWHM$_{\rm PSF}^{\rm tot}$ &
$M_{\rm BH}$ &
$M/L_K$ &
$M/L_K^{0\arcsec}$ &
$M/L_K^{2\arcsec}$ &
$q_{\rm min}$ &
$\sigma_z/\sigma_R$ &
$(\sigma_r/\sigma_\theta)_0$ &
$(\sigma_r/\sigma_\theta)_\infty$ \\
            & ($\arcsec$)& ($\times 10^7$ \Msun) & (\Msun/\Lsun) & (\Msun/\Lsun) & (\Msun/\Lsun) & & \\
(1)         & (2)        & (3)          & (4)           & (5)           & (6)           & (7)   & (8) & (9) & (10) \\
\hline
\multicolumn{10}{c}{\jamcyl\ models with constant anisotropy} \\
\hline
Constant & $0.08$ & $4.52 \pm 0.15$ & $4.45 \pm 0.07$ & \dots & \dots & $0.22 \pm 0.05$ & $0.97 \pm 0.02$ & \dots & \dots \\
Constant & $0.15$ & $4.06 \pm 0.18$ & $4.63 \pm 0.10$ & \dots & \dots & $0.08 \pm 0.05$ & $0.98 \pm 0.02$ & \dots & \dots \\
Constant & $0.20$ & $4.10 \pm 0.14$ & $4.50 \pm 0.09$ & \dots & \dots & $0.25 \pm 0.06$ & $1.00 \pm 0.02$ & \dots & \dots \\
Varying   & $0.08$ & $3.75 \pm 0.19$ & \dots & $5.74 \pm 0.23$ & $3.34 \pm 0.20$ & $0.28 \pm 0.05$ & $0.98 \pm 0.02$ & \dots & \dots \\
Varying   & $0.15$ & $3.36 \pm 0.14$ & \dots & $5.92 \pm 0.21$ & $3.42 \pm 0.16$ & $0.31 \pm 0.05$ & $0.99 \pm 0.02$ & \dots & \dots \\
Varying   & $0.20$ & $3.73 \pm 0.17$ & \dots & $5.54 \pm 0.27$ & $3.75 \pm 0.22$ & $0.36 \pm 0.08$ & $1.03 \pm 0.02$ & \dots & \dots \\
\hline
\multicolumn{10}{c}{\jamsph\ models with radially-varying logistic anisotropy} \\
\hline
Constant & $0.08$ & $4.39 \pm 0.40$ & $3.54 \pm 0.64$ & \dots & \dots & $0.26 \pm 0.04$ & \dots & $0.94 \pm 0.18$ & $1.30 \pm 0.23$ \\
Constant & $0.15$ & $4.27 \pm 0.30$ & $3.56 \pm 0.25$ & \dots & \dots & $0.26 \pm 0.05$ & \dots & $0.78 \pm 0.10$ & $1.30 \pm 0.05$ \\
Constant & $0.20$ & $4.55 \pm 0.44$ & $3.38 \pm 0.57$ & \dots & \dots & $0.19 \pm 0.05$ & \dots & $0.66 \pm 0.16$ & $1.30 \pm 0.23$ \\
Varying  & $0.08$ & $3.70 \pm 0.43$ & \dots & $4.40 \pm 0.97$ & $2.96 \pm 0.48$ & $0.29 \pm 0.25$ & \dots & $1.00 \pm 0.26$ & $1.30 \pm 0.23$ \\
Varying  & $0.15$ & $3.42 \pm 0.45$ & \dots & $4.59 \pm 1.01$ & $3.00 \pm 0.52$ & $0.29 \pm 0.05$ & \dots & $1.00 \pm 0.25$ & $1.29 \pm 0.22$ \\
Varying  & $0.20$ & $4.28 \pm 0.61$ & \dots & $3.76 \pm 0.94$ & $3.17 \pm 0.80$ & $0.20 \pm 0.06$ & \dots & $0.74 \pm 0.34$ & $1.30 \pm 0.26$ \\
\hline
\end{tabular}
\label{tab:adamet}
\noindent\parbox{\linewidth}{\textit{Note:} Column (1): Assumed $M/L_K$ profile type. Column (2): Total FWHM of the synthetic PSF model. Column (3): Best-fit black hole mass. Column (4): Best-fit constant $M/L_K$. Columns (5) and (6): Best-fit central and outer $M/L_K$ for the varying profile. Column (7): Best-fit intrinsic axial ratio. Column (8): Best-fit vertical anisotropy for \jamcyl\ models. Column (9): Best-fit central radial anisotropy for \jamsph\ models. Column (10): Best-fit outer radial anisotropy for \jamsph\ models.}
\end{table*}

We determine the SMBH mass, $M_{\rm BH}$, and other model parameters by performing a Markov Chain Monte Carlo (MCMC) analysis for each JAM model. To explore the parameter space, we employ the adaptive Metropolis algorithm \citep{Haario01} as implemented in the \textsc{adamet} code\footnote{v2.0.9: \url{https://pypi.org/project/adamet/}} \citep{Cappellari2013a}.

For each model, we generate a chain of $10^5$ steps. We discard the initial 20\% as the burn-in phase and use the remaining samples to construct the posterior probability distributions. The most probable parameter values are taken as the maximum of the likelihood, with uncertainties derived from the 1$\sigma$ (16th--84th percentile) and 3$\sigma$ (0.14th--99.86th percentile) confidence intervals of the marginalized posteriors.

Dynamical modeling of numerous, high-precision kinematic data points, as in our case, presents two well-known challenges:
\begin{enumerate}
    \item \textbf{Underestimated Uncertainties:} The formal statistical errors on derived parameters can become unrealistically small, often dwarfed by unmodelled systematic effects.
    \item \textbf{Dominance of Large-Radii Data:} The $\chi^2$ statistic can be disproportionately influenced by the large number of data points at large radii. This risks biasing the $M_{\rm BH}$ measurement, which should be dictated primarily by the kinematics within the black hole's SOI.
\end{enumerate}

To mitigate the first issue, we adopt an error inflation scheme. We base our approach on the heuristic $\Delta\chi^2$-increase method of \citet{vandenBosch2009}, which, in a Bayesian framework, is equivalent to inflating the kinematic measurement errors by a factor of $(2N)^{1/4}$, where $N$ is the number of data points \citep[section~6.1]{Mitzkus2017}. While this technique is now standard for ALMA-based $M_{\rm BH}$ measurements \citep[e.g.,][]{Smith19, Nguyen2020, Nguyen2021, Nguyen2022, Ngo2025a}, a uniform inflation across all radii does not resolve the second challenge.

Therefore, we apply the inflation selectively. We preserve the formal kinematic uncertainties for all data points inside the SOI ($r \leq r_{\rm SOI} \approx 0.45\arcsec$), where the SMBH's gravitational potential dominates. For the $N_{r > r_{\rm SOI}} = 338$ kinematic bins outside this radius, we inflate their associated errors by a factor of $(2N_{r > r_{\rm SOI}})^{1/4}$. This refined strategy ensures the $M_{\rm BH}$ determination is driven by the central kinematics while still accounting for potential systematic errors at larger radii. This selective approach has been successfully employed in previous dynamical studies using integral-field data from Gemini/NIFS \citep{Drehmer2015} and MUSE \citep[section~4.3]{Thater2022}. 

 \begin{figure*}
     \centering
     \includegraphics[height=0.48\textwidth]{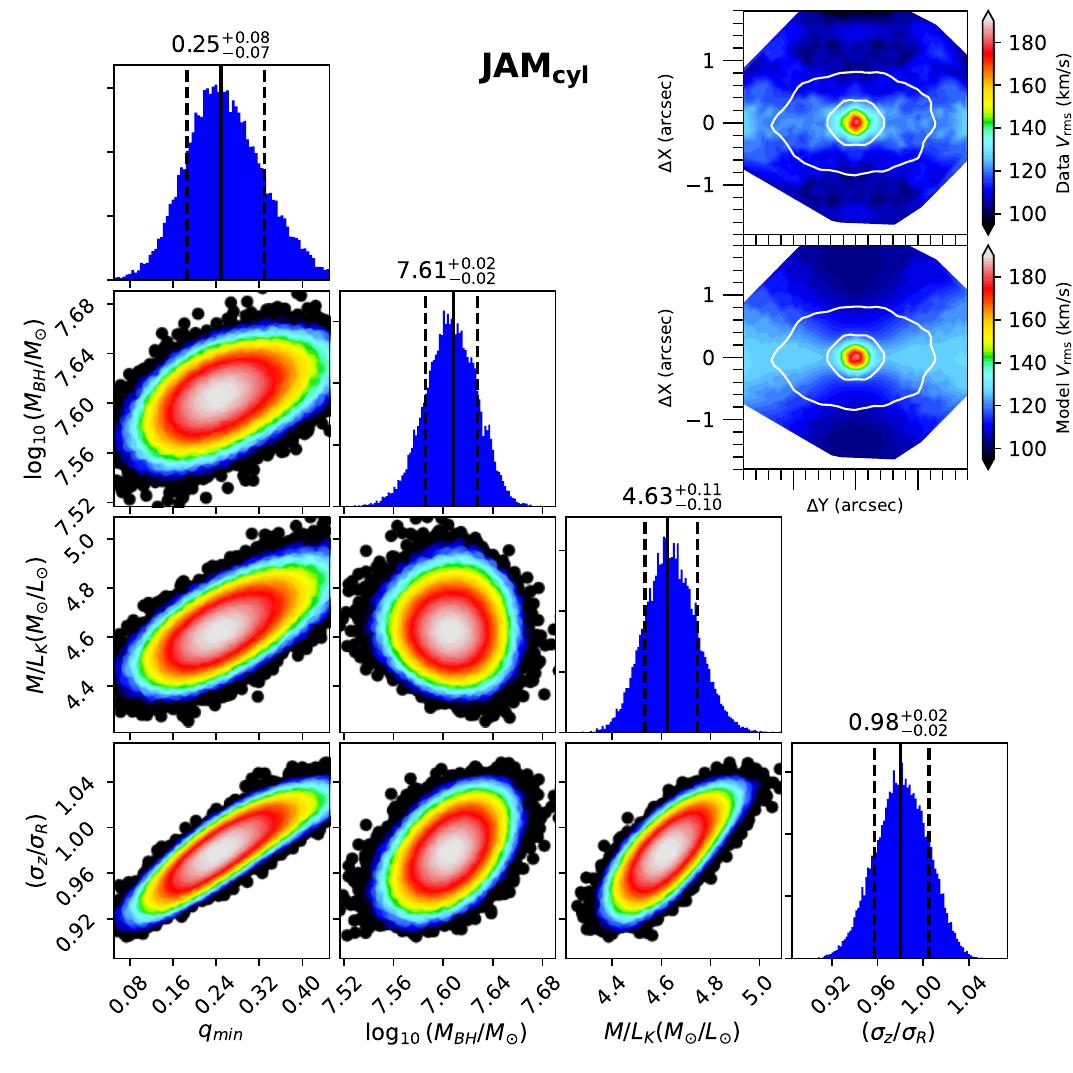}
     \hspace{3mm}
     \includegraphics[height=0.48\textwidth]{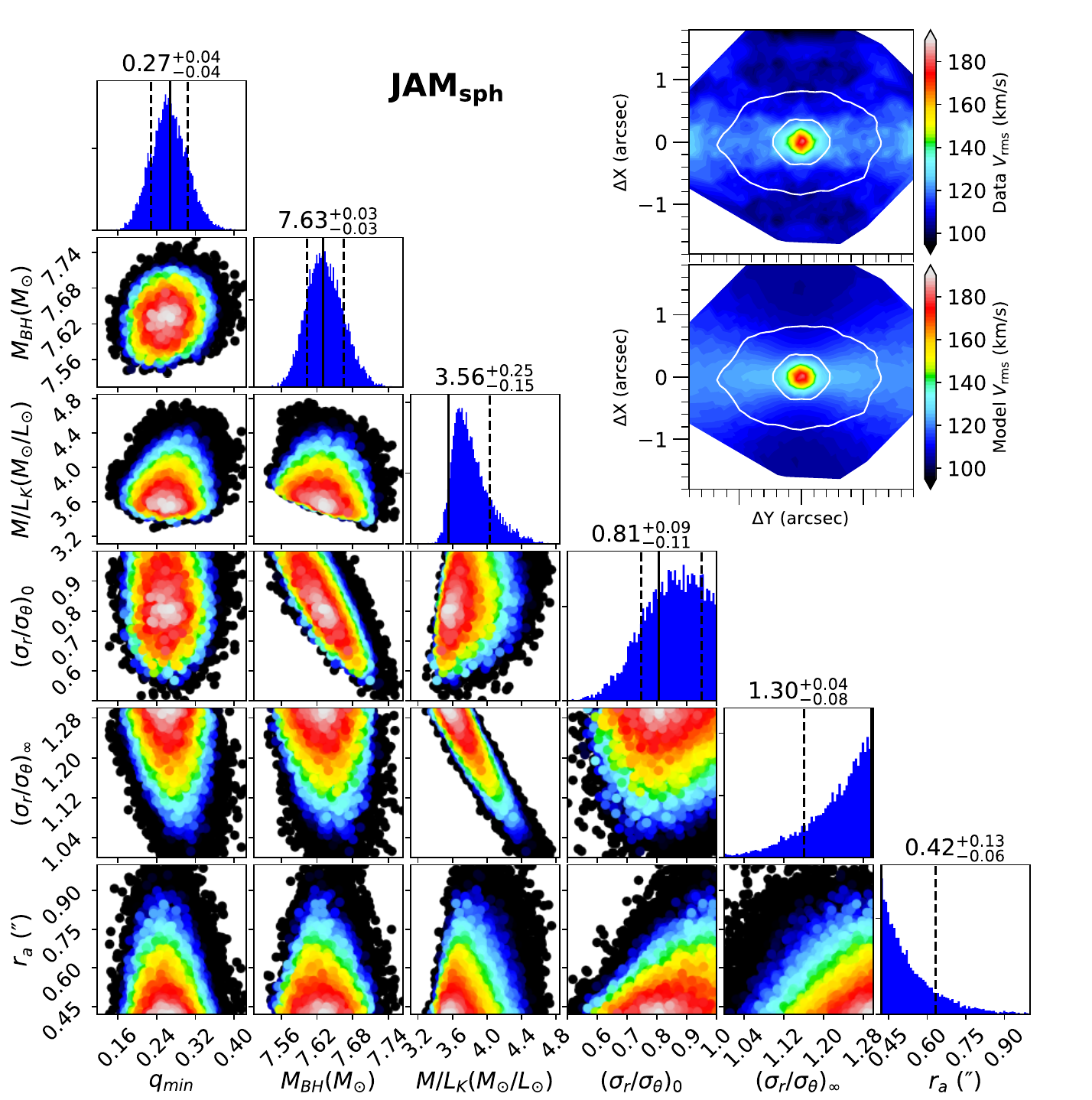}
     \caption{{\it Left:} Posterior distributions for the \jamcyl\ model with constant \ml$_K$ and anisotropy, obtained after the burn-in phase of the \textsc{adamet} MCMC optimization, are shown for the nucleus stellar kinematics of NGC~4258 from the NIRSpec G235H/F170LP data. The triangle plot displays the 2D projected probability distributions of the model parameters ($q_{\rm min}$, $M_{\rm BH}$, $M/L_K$, $\sigma_z/\sigma_R$), with 1D marginalized histograms along the diagonal. Solid vertical lines mark the best-fit values, while dashed lines indicate the 1$\sigma$ uncertainties. Inset panels in the upper right compare the observed and model-predicted $V_{\rm rms}$ using a consistent velocity scale. Both  $V_{\rm rms}$ maps were symmetrized under the assumption of axisymmetry and reoriented with the major and minor axes aligned along the horizontal and vertical directions, respectively, using the \textsc{symmetrize\_velfield} routine from the \textsc{plotbin}\footnote{v3.1.7: \url{https://pypi.org/project/plotbin/}} package. Both the symmetrized data and model prediction were visualized using \textsc{plot\_velfield}. The strong agreement across the NIRSpec FoV confirms the model’s fidelity. {\it Right:} Same as the left panel, but applied the \jamsph\ models with logistic varying $\beta_r$ profile and the model parameters ($q_{\rm min}$, $M_{\rm BH}$, $M/L_K$, $(\sigma_r/\sigma_\theta)_0$, $(\sigma_r/\sigma_\theta)_\infty$, $r_a$).}
     \label{fig:adamet}
 \end{figure*}

\begin{figure}
     \centering
     \includegraphics[width=\linewidth]{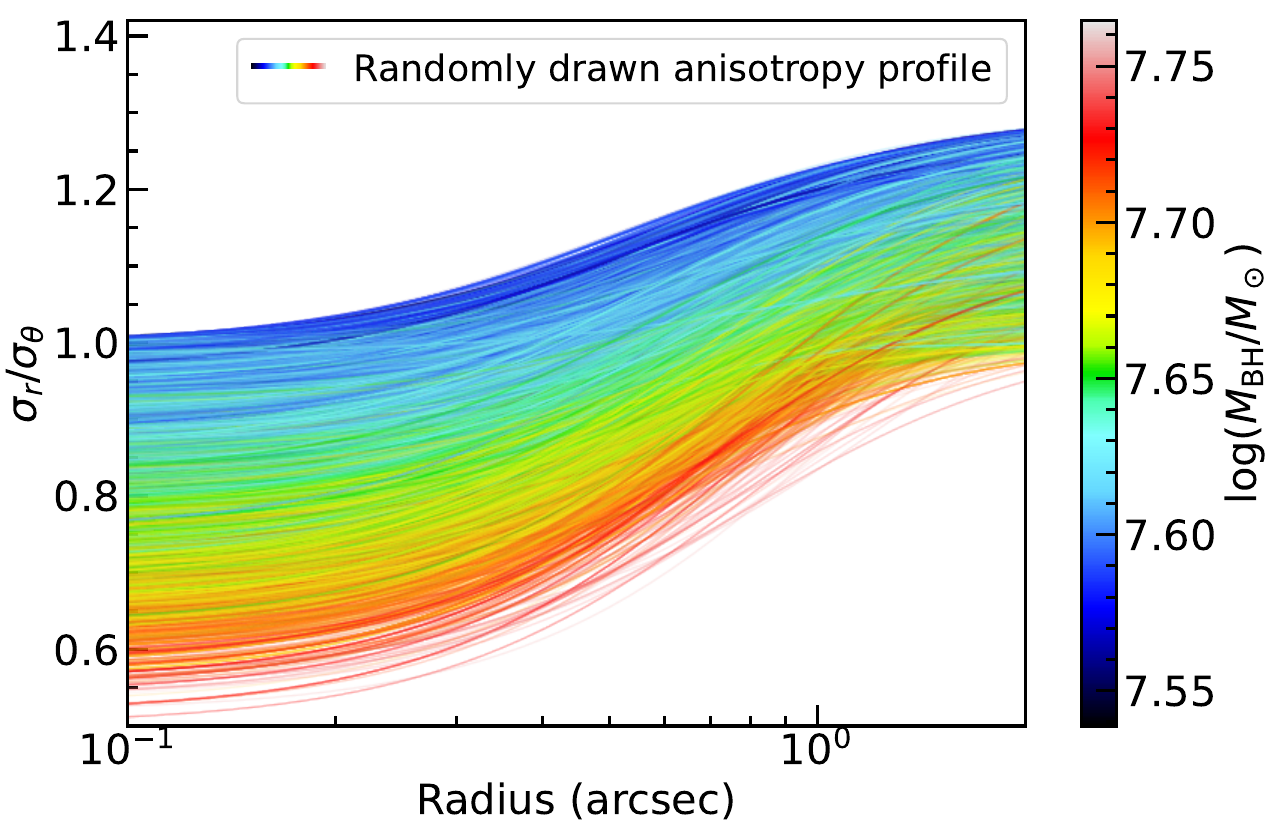}
     \caption{Plot of 100,000 anisotropy profiles randomly sampled from the MCMC chain, color-coded by their corresponding \Mbh\ values. The profiles show clear evidence of a radially increasing anisotropy ratio, though they exhibit significant variation due to the mass–anisotropy degeneracy.}
     \label{fig:adamet_betavary}
 \end{figure}

\section{Results and Discussion}\label{sec:results}

We summarize in \autoref{tab:adamet} the best-fit parameters and their $1\sigma$ uncertainties for the 12 JAM models used to fit the NIRSpec G235H/F170LP $V_{\rm rms}$ data. Across all model assumptions, the inferred \Mbh\ ranges from $(3.36$--$4.55)\times 10^7$~\Msun, and the \ml$_K$ values span $(3.38$--$4.63)$~\Msun/\Lsun. While both \Mbh\ and \ml$_K$ are relatively insensitive to the choice of PSF, they are more strongly affected by the assumed velocity ellipsoid alignment. Specifically, the \jamcyl\ models yield \Mbh\ and \ml$_K$ values that are approximately 23\% higher than those from the \jamsph\ models. In addition, the inferred $q_{\rm min}$ values range from 0.19 to 0.36, corresponding to inclination angles of $i \sim (65$--$73)^\circ$.  In \autoref{fig:summary_all_BH_ngc4258}, we present our 12 \Mbh\ measurements alongside the benchmark water-maser result \citep{Herrnstein1999} and stellar-dynamical estimates from \hst/STIS \citep{Siopis2009} and Gemini/NIFS \citep{Drehmer2015}. Uncertainties are shown as error bars, providing a clear, presentation-ready summary of this benchmark extragalactic constraint on \Mbh\ in NGC~4258.

\autoref{fig:1D-vrms-cyl} presents the $V_{\rm rms}$ profiles of all 12 best-fitting JAM models (with their 1$\sigma$ uncertainties), extracted along the major and minor axis of NGC~4258. These are compared directly to the observed $V_{\rm rms}$ profiles from the NIRSpec G235H/F170LP data, extracted in the same way. All models generally provide a good match to the data and to one another across the NIRSpec FoV, despite differing assumptions of the \ml$_K$ profile and orbital anisotropy. A systematic offset is seen at the very center, corresponding to the innermost kinematic bins, likely due to differences in the treatment of our synthetic NIRSpec PSF in the modeling (\autoref{sec:nirspec_psf} and \autoref{tab:psf}). In particular, models assuming PSF 3 (FWHM$_{\rm PSF, tot}\approx0\farcs2$) fail to reproduce the central $V_{\rm rms}$ values, missing by $\approx$20 \kms\ (see right-hand column of \autoref{fig:1D-vrms-cyl}). In contrast, the best-fitting models using PSF 1 (FWHM$_{\rm PSF, tot}\approx0\farcs08$; left-hand column) and PSF 2 (FWHM$_{\rm PSF, tot}\approx0\farcs15$; middle column) yield a significantly better fit to the G235H/F170LP data. Although the JAM models using our synthetic PSF 1 provide the best match to the $V_{\rm rms}$ data, including the innermost bins, those based on synthetic PSF 2 match the central data within the $1\sigma$ $V_{\rm rms}$ uncertainties. We prefer the best-fit parameters from the models using synthetic PSF 2, as its total FWHM closely matches that of the empirically determined PSF from \citet{D’Eugenio2024NaAs, D'Eugenio2025}.

Our 12 JAM runs provide an ensemble of broadly acceptable models that span a range of assumptions and explore different systematic effects. This approach allows us to move beyond the common practice of relying on the unreliable formal uncertainties from a single `preferred' model and instead compute robust ensemble statistics. The mean black hole mass from our 12 models is $(4.01 \pm 0.12) \times 10^7$~\Msun, where the uncertainty is the standard error of the mean. However, the mean relies on the assumption that the distribution of model results is Gaussian, which may not be the case, and its uncertainty may be unrealistically small. We therefore also compute a more robust estimate: the median with its bootstrap 68\% (1$\sigma$) confidence intervals, which yields $M_{\rm BH} = (4.08^{+0.19}_{-0.33}) \times 10^7$~\Msun. We adopt this median value as our final, more robust estimate, in part because of its slightly larger and more conservative uncertainty. Both the mean and median values are consistent with the benchmark maser value within their uncertainties. Our adopted mass is just 5\% larger than the maser-based value of $(3.9 \pm 0.1) \times 10^7$~\Msun\ \citep{Herrnstein1999}, 15\% smaller than the previous JAM stellar-dynamical estimate of $(4.8\pm0.3) \times 10^7$~\Msun\ from \citet[converted to $1\sigma$ confidence]{Drehmer2015}, and 24\% larger than the Schwarzschild-models value of $(3.3 \pm 0.2) \times 10^7$~\Msun\ from \citet{Siopis2009}.

The best-fit \jamcyl\ models are generally isotropic, with constrained values of $\sigma_z/\sigma_R\sim1$ (i.e., $\beta_z\sim0$), regardless of whether a constant or varying \ml$_K$ is assumed. In contrast, the derived \Mbh\ shows a stronger dependence on the assumed \ml$_K$ profile in the nucleus of NGC~4258: \Mbh\ values are (6--18)\% higher or (2--12)\% lower than the maser-based \Mbh\ \citep{Herrnstein1999} under constant or linearly varying \ml$_K$ assumptions, respectively. These smaller \Mbh\ values, constrained from the linearly varying \ml$_K$ profile, result from its negative slope. A similar trend in the dependence of the constrained \Mbh\ on the assumed \ml$_K$ profile is also observed in the best-fit \jamsph\ models with logistic anisotropy. This result suggests the possible presence of dust within the NIRSpec IFU FoV of NGC~4258, contributing non-thermal light similar to AGN emission, but at larger radii beyond the galaxy center. This supports our inference of a spatially varying stellar light fraction, $\gamma < 1$, especially elongated along the minor axis as shown in panel F of \autoref{fig:kin-map}.

The left panel of \autoref{fig:adamet} shows an example of the 2D posterior distributions for each model parameter of the \jamcyl\ model with constant \ml$_K$, constant anisotropy, and PSF 2. Point colors indicate the relative likelihood (white denotes the maximum likelihood and 1$\sigma$, and black marks the 3$\sigma$ confidence levels, CLs). The accompanying 1D histograms are used to derive the best-fit values and 1$\sigma$ uncertainties, which reflect the propagation of stellar kinematics and random errors. All four parameters are well constrained. The figure also illustrates the agreement between the observed $V_{\rm rms}$ and that predicted by the best-fit \jamcyl\ model.    

\begin{figure}
    \centering
    \includegraphics[width=0.98\linewidth]{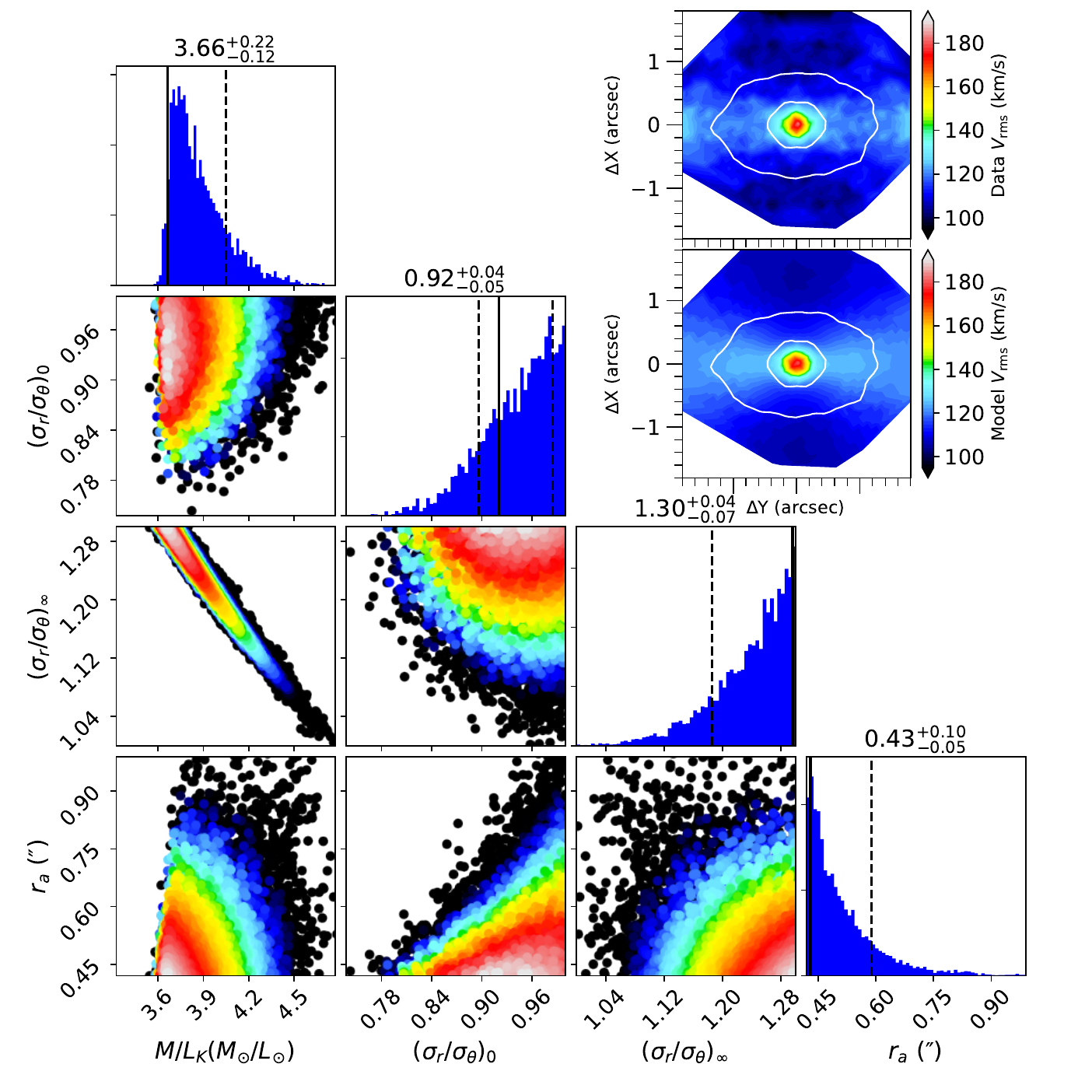}
    \caption{Same as \autoref{fig:adamet}, but for the \jamsph\ models with the $M_{\rm BH}$ fixed to the maser–based estimate from \citet{Herrnstein1999} and the inclination fixed to that of the best-fitting \jamsph\ model.} 
        \label{fig:water-maser-BH}
\end{figure}

Similarly, in the right panel of \autoref{fig:adamet}, we show the 2D posterior distributions for each model parameter of the \jamsph\ model with constant \ml$_K$, logistic anisotropy, and PSF 2.   The posterior distributions for the JAM$_{\rm sph}$ model appear to hit the boundary set by our anisotropy prior, indicating that certain parameters are only weakly constrained by the data alone and are partially driven by the priors. We emphasize that this behavior is expected and, in fact, necessary. In galaxy dynamics analyses of single galaxies, observational data are often insufficient to uniquely constrain all model parameters due to intrinsic degeneracies, such as the well-known mass-anisotropy degeneracy \citep{Binney1982}. If the posterior distributions were entirely unaffected by the priors, it would imply that the priors were unnecessary. Instead, the priors serve their intended purpose: to incorporate external physical knowledge---accumulated over decades of dynamical modeling studies---to break degeneracies that the data cannot resolve on their own. Often, priors are assumed to be Gaussian, and their influence on the posterior is not as obvious as when priors are set as sharp boundaries, as we do here. However, in either case, priors affect the posterior, and the ability to enforce them is an integral feature and a key strength of the Bayesian approach. The JAM method is particularly well-suited to enforcing priors, which have always been a key feature of its applications. This approach is superior to assuming complete ignorance (flat priors), which can lead to unphysical solutions, as historically demonstrated by the extreme anisotropy required to explain M87 kinematics without a black hole \citep{Binney1982}. By leveraging informative priors, we ensure that our $M_{\rm BH}$ estimates remain robust and physically consistent with our current understanding of galaxy structure.

\autoref{fig:adamet_betavary} illustrates a set of 100,000 anisotropy profiles randomly drawn from the 2D posterior distributions in the right panel of \autoref{fig:adamet}. These profiles exhibit a radially increasing trend, often found in galaxy centers \citep[fig.~10]{Thomas2014, Cappellari2026}. Notably, the profiles tend to transition from roughly constant values at larger radii to lower values near $\approx$0\farcs42–0\farcs6. This transition scale is close to the BH’s SOI in NGC~4258, where stellar random motions are strongly enhanced by the BH’s gravitational potential, resulting in tangential anisotropy. Beyond the SOI, the anisotropy gradually shifts toward radially dominated stellar motions. We observe a strong \Mbh–anisotropy degeneracy, in which higher \Mbh\ values correspond to more tangential velocity anisotropy, while lower \Mbh\ values correspond to more radial anisotropy, consistent with that observed in the core elliptical galaxy M87 \citep[figure 15 of][]{Simon2024}. 

Within the 3$\sigma$ CL region of the 2D posterior distributions, we find no significant covariance between \Mbh\ and \ml$_K$ in the JAM models, indicating that both parameters are well constrained. Additionally, a positive covariance is observed between \ml$_K$ and the orbital anisotropy parameter (i.e., $\sigma_\theta/\sigma_r$ in the \jamsph\ model and $\sigma_z/\sigma_R$ in the \jamcyl\ model), suggesting that variations in anisotropy can partially compensate for changes in \ml$_K$ during the fitting process.

Given that the water maser–based \Mbh\ measurement from \citet{Herrnstein1999} remains the most accurate extragalactic determination, we tested a \jamsph\ model with constant \ml$_K$, logistic anisotropy, and PSF 2, in which \Mbh\ was fixed to this value and $q_{\rm min}$ was set to that of the best-fitting \jamsph\ model, while allowing \ml$_K$, $r_a$, and the $(\sigma_r/\sigma_\theta)_0$ and $(\sigma_r/\sigma_\theta)_{\infty}$ ratios to vary freely. As shown in \autoref{fig:water-maser-BH}, we find a strong anti-covariance between \ml$_K$ and $(\sigma_r/\sigma_\theta)_{\infty}$, reflecting the \Mbh–anisotropy degeneracy. The best-fit \ml$_K$ remains fully consistent with that of the original \jamsph\ model. For smaller black hole masses, the anisotropy becomes less tangential and more radial, with $(\sigma_r/\sigma_\theta)_0$ increasing by 12\%.

Our stellar-based \Mbh\ constraints, derived under both cylindrical and spherical velocity ellipsoid assumptions, yield $r_{\rm SOI} \approx 0\farcs45$ (or 16 pc). These radii are $\approx$two–three times larger than the spatial resolution of the \jwst\ observations, demonstrating that our \Mbh\ estimates are robust and spatially well resolved. The kinematic signature of the SMBH is clearly detected within the central $\approx$55 spaxels.

\section{Conclusions}\label{sec:conclusion}

We have measured the SMBH mass in the benchmark galaxy NGC~4258 using high-resolution 2D stellar kinematics from \jwst/NIRSpec IFU observations. By applying JAM models to the stellar velocity distribution derived from the CO bandheads, we robustly determined the central dark mass. Our main findings are summarized as follows:

\begin{enumerate}

    \item We successfully recovered the stellar kinematics by separating the stellar light from the non-thermal AGN continuum, which contributes up to 56\% of the flux in the central spaxel. This was achieved by fitting a fixed optimal stellar template combined with additive and multiplicative polynomials to model the AGN contribution.

    \item The resulting kinematic maps show clear rotation ($\pm$80~\kms) and a sharp increase in velocity dispersion towards the center, peaking at $191 \pm 12$~\kms. This central kinematic spike is a clear signature of the SMBH's gravitational influence and is resolved by the NIRSpec data.

    \item We ran a grid of 12 JAM models to explore systematic uncertainties from the PSF, M/L profile, and orbital anisotropy. The ensemble median and corresponding bootstrap 68\% (1$\sigma$) confidence interval of these models yields a black hole mass of $M_{\rm BH} =(4.08^{+0.19}_{-0.33}) \times 10^7$~\Msun. This result, which incorporates both statistical and systematic uncertainties, is in excellent agreement with the precise maser-based mass.

    \item Our analysis shows that the choice of PSF significantly impacts the fit quality. Models using a synthetic PSF with a FWHM of 0\farcs15 provide the best match to the central kinematics, consistent with recent empirical PSF measurements for NIRSpec.

    \item This work demonstrates that \jwst/NIRSpec stellar kinematics can deliver precise and accurate SMBH masses, even in AGN-dominated nuclei. Our result for NGC~4258 validates the stellar dynamical method against the gold-standard maser technique, paving the way for robust mass measurements in more distant galaxies.

\end{enumerate}

\section*{Acknowledgements}
The authors would like to thank the anonymous referee for their careful reading and useful comments, which helped to improve the paper greatly.  Research conducted by T.N.L. is partially supported by a grant from the Simons Foundation to IFIRSE, ICISE (916424, N.H.).  N.T. would like to acknowledge partial support from the UKRI grant ST/X002322/1 for UK ELT Instrument Development at Oxford. M.P. acknowledges support through the grants PID2021-127718NB-I00 and RYC2023-044853-I, funded by the Spanish Ministry of Science and Innovation/State Agency of Research MCIN/AEI/10.13039/501100011033 and El Fondo Social Europeo Plus FSE+. M.P.S. acknowledges support under grants RYC2021-033094-I, CNS2023-145506, and PID2023-146667NB-I00 funded by MCIN/AEI/10.13039/501100011033 and the European Union NextGenerationEU/PRTR. 

This research is based on observations made with the NASA/ESA Hubble Space Telescope obtained from the Space Telescope Science Institute, which is operated by the Association of Universities for Research in Astronomy, Inc., under NASA contract NAS 5–26555. This work is based [in part] on observations made with the NASA/ESA/CSA James Webb Space Telescope. The data were obtained from the Mikulski Archive for Space Telescopes at the Space Telescope Science Institute, which is operated by the Association of Universities for Research in Astronomy, Inc., under NASA contract NAS 5-03127 for JWST. These observations are associated with program \#02016.

Some/all of the data presented in this article were obtained from the Mikulski Archive for Space Telescopes (MAST) at the Space Telescope Science Institute. The specific observations analyzed can be accessed via \dataset[doi:10.17909/baxx-nf29]{https://doi.org/10.17909/baxx-nf29}.

\facility{\jwst/NIRSpec.}

\software{{\tt Python~3.12:} \citep{VanRossum2009}, 
{\tt Matplotlib~3.6:} \citep{Hunter2007}, 
{\tt NumPy~1.22:} \citep{Harris2020}, 
{\tt SciPy~1.3:} \citep{Virtanen2020},  
{\tt photutils~0.7:} \citep{bradley2024}, 
{\tt AstroPy~5.1} \citep{AstropyCollaboration2022}, 
{\tt AdaMet 2.0} \citep{Cappellari2013a}, 
{\tt JamPy~7.2} \citep{Cappellari2020}, 
{\tt pPXF~8.2} \citep{Cappellari2023}, 
{\tt vorbin~3.1} \citep{Cappellari2003}, and
{\tt MgeFit~5.0} \citep{Cappellari2002}.
}

\bibliographystyle{aasjournalv7}
\bibliography{references} 

\end{document}